\documentclass[aps,pra,reprint,superscriptaddress,amsmath,showpacs,amsfonts,amssymb,longbibliography]{revtex4-1}

\usepackage{graphicx}
\usepackage{bm}
\usepackage{color}
\usepackage{MnSymbol}
\usepackage{enumerate}
\usepackage{bbold}
\usepackage{subfigure}
\usepackage{units}
\usepackage[usenames,dvipsnames]{xcolor}
\usepackage[colorlinks=true,linkcolor=MidnightBlue,citecolor=blue,filecolor=green,urlcolor=PineGreen]{hyperref}

\def\la{\langle}
\def\ra{\rangle}

\def\be{\begin{equation}}
\def\ee{\end{equation}}

\begin{document}

\title{Gravitational sensing with weak value based optical sensors}
\author{Andrew N. Jordan}
\affiliation{Department of Physics and Astronomy, University of Rochester, Rochester, New York 14627, USA}
\affiliation{Institute for Quantum Studies, Chapman University, Orange, CA 92866, USA}
\affiliation{A. N. Jordan Scientific, LLC, 91 Westerloe Ave., Rochester, NY 14620, USA}
\author{Philippe Lewalle}
\affiliation{Department of Physics and Astronomy, University of Rochester, Rochester, New York 14627, USA}
\affiliation{A. N. Jordan Scientific, LLC, 91 Westerloe Ave., Rochester, NY 14620, USA}
\author{Jeff Tollaksen}
\affiliation{Institute for Quantum Studies, Chapman University, Orange, CA 92866, USA}
\affiliation{Schmid College of Science and Technology, Chapman University, Orange, CA 92866, USA}
\author{John C. Howell}
\affiliation{Racah Institute of Physics, The Hebrew University of Jerusalem, Jerusalem, Israel, 91904}
\affiliation{Department of Physics and Astronomy, University of Rochester, Rochester, New York 14627, USA}
\begin{abstract}
Using weak values amplification angular resolution limits, we theoretically investigate the gravitational sensing of objects. By inserting a force-sensing pendulum into a weak values interferometer, the optical response can sense accelerations to a few 10's of \unit[]{zepto-g Hz$^{-1/2}$}, with optical powers of \unit[1]{mW}.  We convert this precision into range and mass sensitivity, focusing in detail on simple and torsion pendula.  Various noise sources present are discussed, as well as the necessary cooling that should be applied to reach the desired levels of precision.
\end{abstract}

\date{\today}
\maketitle

\section{Introduction}
We explore fundamental limits in precision gravimetry using weak value amplification techniques \cite{Aharonov88,Ritchie91,Hosten08,Dixon09,Brunner10,Dressel2013,Lyons2015,Martinez17}. Weak values were born through asking fundamental questions about quantum measurement limits \cite{Aharonov88}. Unlike expectation values, weak values consider a normalized expectation of an operator (e.g., the Pauli operator $\hat{A} = |+\ra\la +| - |-\ra\la -|$) using pre- and post-selected quantum states $\psi_{i,f}$ 
\be
A_w = \frac{\la \psi_f | A|\psi_i\ra}{\la \psi_f |\psi_i\ra}.
\ee
Because weak values can be much larger than their respective expectation values when $\la \psi_f |\psi_i\ra \rightarrow 0$, they have been used to amplify small effects.  Weak value amplification has been shown to be exceptionally valuable in suppressing technical noise in precision measurements \cite{Starling09,Feizpour11,Jordan14,Viza15,pang2016protecting,sinclair2017weak,Lyons2017}. While these techniques do not beat the shot noise limit (with some exceptions, see e.g.~Ref.~\cite{jordan2015heisenberg}), they can come close to reaching it because of the dramatically suppressed technical noise.  Of particular interest is the recent inverse weak value work where an angular tilt measurement noise floor of 200 frad Hz$^{-1/2}$ was achieved.  Remarkably, this sensitivity was for signals down to 1 Hz \cite{Martinez17}, where noise suppression can be incredibly difficult.  This tilt corresponds to a displacement of less than a hair's breadth at the distance of the moon \cite{Steinberg10} in one second of measurement time using only a few milliwatts of laser power.  We show that if these techniques can be used, even at the classical optical fundamental limits, for precision gravimetry, they would push gravimetric sensitivity by several orders of magnitude beyond the state-of-the-art.  

Precision gravimetry \cite{Wahr04,Bingham10,Bell98,Leeuwen00,Diorio03,Romaides01,Peters01,Luther82,Kuroda95,Karagioz96,Bagley97,Gundlach00,Quinn01,Armstrong03,Kleinevoss99,Parks10,Peters99,Mcguirk02,Dimopoulos07,Lamporesi08,Sorrentino10,Rosi14,Goodkind99} is used extensively in mapping the earth's local gravity \cite{Wahr04,Bingham10}, oil and gas exploration \cite{Bell98}, mining \cite{Leeuwen00}, mapping temporal geological shifts, the determination of Newton's gravitational constant \cite{Luther82, Kuroda95, Karagioz96, Bagley97, Gundlach00, Quinn01, Armstrong03, Kleinevoss99, Parks10, Peters99, Mcguirk02, Dimopoulos07, Lamporesi08, Sorrentino10, Rosi14} and gravitationally imaging opaque systems. 
Precision of the order of 1 $\mu$g (1 g = \unit[9.8]{m s$^{-2}$} to 1 nano-g) are often used for mapping geological variations. 
Both relative and absolute measurements are employed. 
A standard in the industry for absolute gravimetry is measuring interference fringes due to the free-fall of a corner cube in one arm of a Mach-Zehnder interferometer with a sensitivity of \unit[100]{nano-g Hz$^{-1/2}$}.   Another competing gravimetric technology employs atomic interferometry achieving a resolution of 100 pico-g after two days of integration \cite{Peters01}. 
The field standard is a superconducting sphere suspended in the field of a superconducting coil achieving 3 pico-g resolution \cite{Goodkind99} after one month integration and 1 pico-g  after one year. 
The most sensitive device to date is Kasevich's 10 m atom interferometer which achieves 500 femto-g after one hour of integration \cite{PhysRevA.91.033629,kasevich2014prospects}. 


The purpose of this paper is to advance a gravitational sensor, whose readout is entirely optical.  
The sensor is a relative gravity sensor, able to sense changes in gravitational fields around it. 
Our design is based around mechanical elements, such as simple and torsion pendula, that are incorporated into an optical interferometer. 
Similar ideas have been recently and independently explored in Ref.~\cite{turner2018development,Ciani2017}. 
This interferometer is constructed to realize the inverse weak value effect, where a continuous optical phase can be read out via a slight change on intensity detectors, typically a split--detector for the discussions in this paper. 
Therefore, we require a gravitational force to cause a change in optical phase. 
This is implemented with a mirror attached to the mechanical element which is suspended. 
When the element undergoes a slight acceleration from the gravitational force, the mirror undergoes a slight tilt, which is the mechanical change which is optically read out. 
Once the device is realized, we find excellent force sensing abilities, due in large part to the extreme sensitivity of the interferometer to optical phase shifts.

The paper is organized as follows: In Sec.~\ref{iwvi}, we discuss the inverse weak value interferometery approach to measuring optical phase shifts.  Interferometer design is given, and we introduce the modifications necessary to incorporate the gravitational sensor as controlling one of the interferometer mirrors.  In Sec.~\ref{gtc}, the design of the mechanical element is discussed, and how the gravitational response of the pendulum can be dynamically sensed.  Sec.~\ref{noise} discusses various noise sources that will be acting on the pendulum, which will mask the underlying gravity signal the detector is sensing.  Ways to mitigate those noise sources are discussed.  Fundamental resolution limits on sensed mass and range of target as calculated in Sec.~\ref{resolution}.  We conclude in Sec.~\ref{conc}.

\section{Inverse weak value interferometry} \label{iwvi}
A specialized weak values interferometer employs a laser beam of transverse width $\sigma$ in a Sagnac interferometer \cite{Dixon09}.   The laser beam enters a beamsplitter and propagates in opposite directions around a Sagnac interferometer.  When the beams recombine, a small relative phase $\phi$ between the two returning beams causes a small transverse tilt, $k$, of the mirror attached to the pendulum to be amplified in the (nearly) dark port of the beamsplitter.  The weak value limit occurs when $\phi \gg k\sigma$.  For this particular setup, the amplification of the small transverse tilt $k$ shows up in the dark port beam as a spatial shift by $k \sigma^2/\phi$.  In terms of weak values, the pre-selected state $\psi_i$ is the field after passing through beamsplitter the first time.  The  post-selected state $\psi_f$ is set by a combination of a phase shift and the second pass through the beamsplitter.  The tilt of the mirror yields a small amount of which-path information of a photon in the interferometer, which is a weak measurement of $\sigma_z$.  

Conversely, for an {\em inverse} weak values experiment, the parameters satisfy the inequality $\phi \ll k\sigma$.  In this case, we fix transverse tilt $k$ and use the known $k$ to amplify a small unknown phase. In this latter experimental regime, the interference pattern of the two beams in the dark port is now a bimodal distribution with a dark fringe at the center of the interference pattern for $\phi=0$.  The dark fringe moves rapidly with small changes in relative phase.  These phase shifts are determined by measuring the relative intensity of the left versus right side of the interference pattern via a split detector. The amplification of the {\em phase}  in this inverse weak value regime is given by the mean shift of the beam in the dark port, $\phi/k$, which now is proportional to the inverse weak value $A_w^{-1}$ \cite{starling2010continuous}.  

\begin{figure}[tb!]
   \centering   \includegraphics[width=0.90\linewidth]{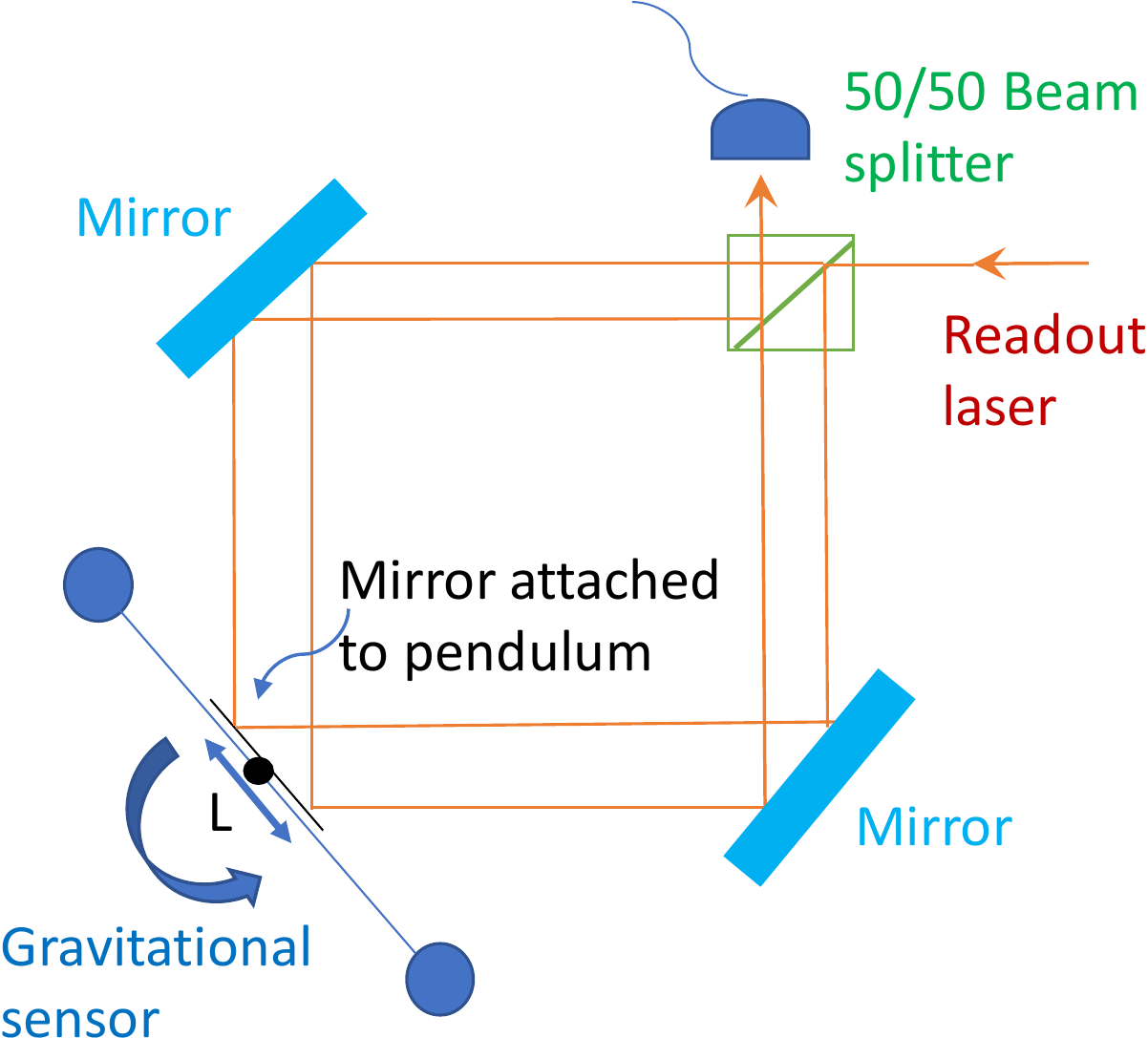}
   \caption{A Sagnac interferometer with a torsion pendulum integrated as a gravity sensor. This cartoon illustrates the type of device considered throughout the manuscript.}
   \hfill\null
   \label{model}
\end{figure}

In Ref.~\cite{Martinez17}, a displaced Sagnac interferometer was used to measure the relative phase shift $\phi$ for this inverse weak value regime.  In a displaced Sagnac interferometer, two beams propagate with a transverse displacement (albeit parallel) in opposite directions.  A small tilt of a mirror inside the interferometer causes a relative phase between the two paths since the path length increases for one path and decreases for the other.  At the output port of the interferometer, the beams are brought back together to interfere with each other. 

The shot noise limited angular resolution can be understood from a geometric argument. The relative phase
between the two paths goes as 
$\delta \phi = 2 \sqrt{2} \pi L \theta/\lambda$, where
$\sqrt{2}$ comes from impinging at 45 degrees, $L$ is the distance
between the centers of the beams propagating in opposite directions (see Fig.~\ref{model}), $\lambda$ is the wavelength of the laser light, and $\theta$ is the tilt angle of the mirror that we are interested in determining.  Assuming the phase can be determined with shot noise limited sensitivity $\Delta \phi=\frac{1}{2\sqrt{N}}$, we find
\begin{equation}
\Delta\theta_{SN}=\frac{1}{4\sqrt{2}\pi}\frac{\lambda}{L\sqrt{N}}, \label{res-limit}
\end{equation}
where $N$ is the number of detected photons.  Using $L=\unit[1]{cm}$, approximately \unit[3]{mW} of laser power, and a wavelength of \unit[500]{nm}, we achieve a shot noise limited angular sensitivity of \unit[30]{frad/Hz$^{1/2}$}. The inverse weak value method of readout for the optical phase $\phi$ can achieve this shot-noise limited sensitivity, up to a factor of $\sqrt{\pi}/2$ associated with the resolution loss on the split detector \cite{knee2014amplification,Jordan14}.

In this work, we use this same inverse weak value setup with a displaced Sagnac.  We consider the physical limitations and sensing capabilities when the tilt mirror in \cite{Martinez17} is replaced with a mirror rigidly connected to a pendulum as shown in Fig. \ref{model}.

\section{Geometry and torque calculation}\label{gtc}

\begin{figure}[tb!]
\begin{centering}
\includegraphics[width=.9\columnwidth, trim={60pt 100pt 200pt 0pt},clip]{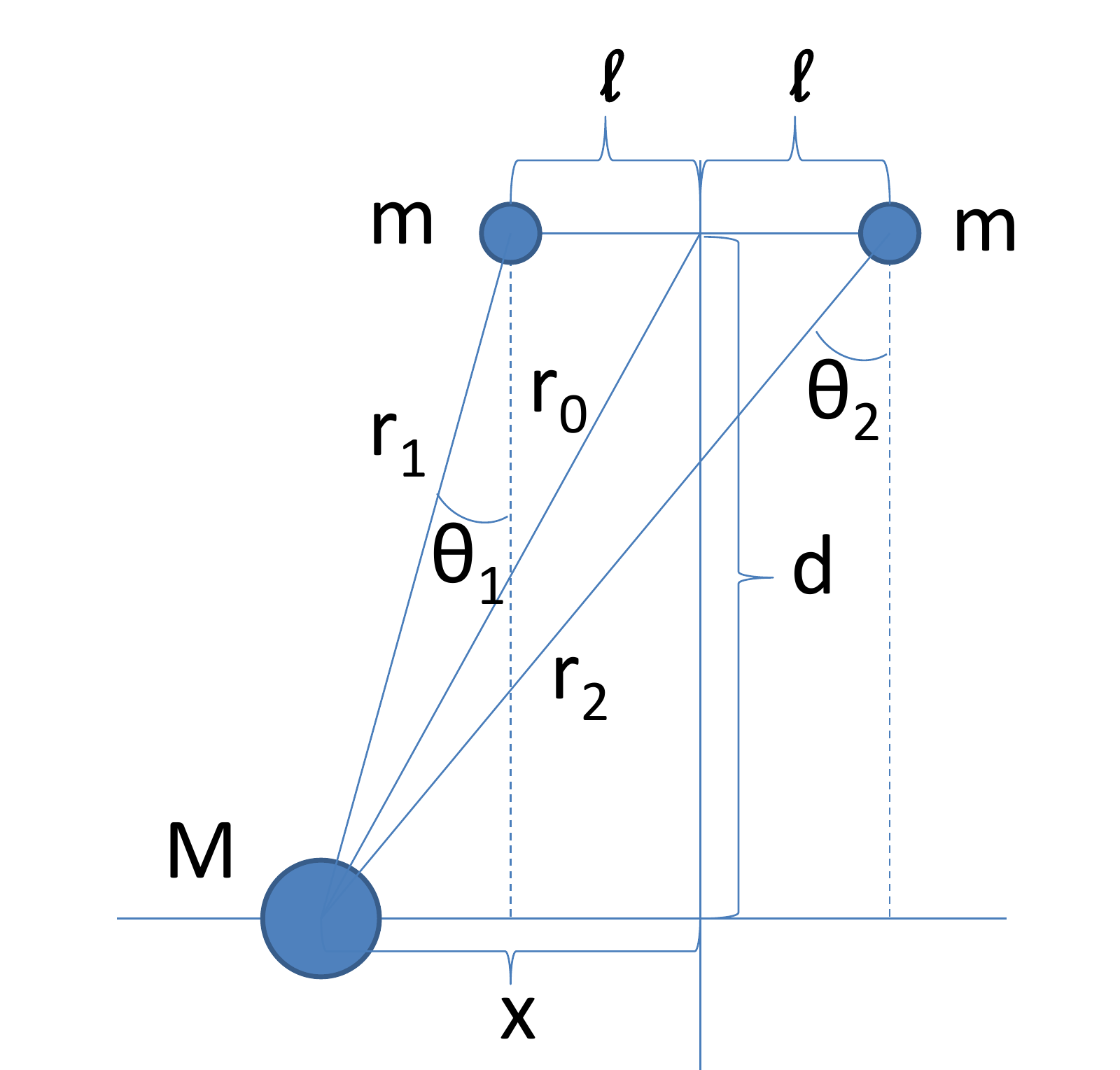}
\caption{A torsion pendulum is formed by attaching two masses $m$ with a rigid, massless rod, of total length $2 \ell$.  We fix the center of mass of the pendulum in one place, allowing it to rotate only in the plane of the figure.  A nearby mass $M$ creates a torque on this torsion pendulum, causing it to rotate to an equilibrium angle $\delta \theta_0$, which is detected optically.  Lengths, angles, and mass labels are shown in the figure. }
\label{fig-geometry-pendulum}
\end{centering}
\end{figure}

In the following, we consider a gravity torsion pendulum with an optical readout via the inverse weak value method \cite{Dixon09} as discussed in the previous section.
Suppose there is a system consisting of a mass $M$ that is detected via a torsion pendulum consisting of two masses $m$, connected via a rigid massless rod of length $2 \ell$.   The mass $M$ is located according to Fig.~\ref{fig-geometry-pendulum} in relation to the oriented torsion pendulum. We will assume $M$ is a point mass for the time being.  To start, let us suppose that the motion of all those objects will be characterized by their moment of inertia $I$ about the axis defined by the pivot.  By way of example, that axis could be comprised of a wire attached to the rigid body consisting of the masses $m$ at both ends.  An external torque exerted from the gravitational force of a massive object will disturb the equilibrium position of the oscillator, which will oscillate until its damps to the new equilibrium position, as described in the next subsection.

\subsection{Converting torque into angle}
In order to detect this small torque, we first recall that the pendula have a linear restoring torque quantified by the torsion spring constant $\kappa$, such that 
\be
\tau_{ext} = - \kappa \delta \theta,
\ee
where $\delta \theta$ denotes the angular distance from its equilibrium position.  Note that we can empirically find $\kappa$ by finding the period of the oscillations.  When the pendulum swings freely, $\tau_{ext} = I \alpha = 2 m \ell^2 {\ddot \theta} = - \kappa \delta \theta$, where $I = 2 m \ell^2$ is the moment of inertia.  Putting this equation in the form $ {\ddot \theta} + \omega_0^2 \theta=0$, the natural frequency of the pendulum is $\omega_0 = (1/\ell)\sqrt{\kappa/(2 m)}$, which may be inverted to find $\kappa$ from a measurement of the frequency, or period of the pendulum ${\cal T}$,
\be
\kappa = \frac{8 \pi^2 m \ell^2}{{\cal T}^2}.\label{kappa}
\ee
Adding in damping of the pendulum brings the dynamics into the form of Eq.~(\ref{theq}).

For a vertical simple pendulum subject to Earth's gravitational field, the restoring force is simply the gravitational acceleration from the earth, causing a restoring torque of $\tau = - g m \ell \delta \theta$ for small angles.  This also gives rise to dynamics of the form Eq.~(\ref{theq}) but with a natural frequency  given by $\omega_0 = \sqrt{g/\ell}$.

The source of the gravitational signal is a mass $M$ near the pendulum, thereby applying an external torque $\tau$.  We wish to measure this signal. We assume that $\tau$ can be time--dependent in general.
Damping of the oscillations will be critical for a quickly responding detector, so we also add a velocity-dependent damping term, to find the equation of motion of a damped/driven oscillator,
\be \label{theq}
\ddot{\theta} + 2 \zeta \omega_0 \dot{\theta} + \omega_0^2 \theta = \tau(t)/I.
\ee
Here, $\omega_0 = \sqrt{\kappa/I}$, and $\zeta$ is a dimensionless damping coefficient. The underdamped case corresponds to $ 0 \leq \zeta < 1$, whereas the overdamped case corresponds to $\zeta > 1$. The general solution for $\tau = 0$ (the homogeneous solution) can be expressed as
\be \begin{split}
\theta_{hom}(t) = &e^{-\zeta \omega_0 t} \bigg \lbrace \theta_0 \left[ e^{\omega_0 t \sqrt{\zeta^2-1}} -\sinh\left(\omega_0 t \sqrt{\zeta^2-1}\right) \right]  \\ &+ \left[ \frac{\zeta \theta_0 + \dot{\theta}_0/\omega_0}{\sqrt{\zeta^2-1}} \right] \sinh\left(\omega_0 t \sqrt{\zeta^2-1}\right) \bigg\rbrace,  \label{homosol}
\end{split}\ee
where $\theta_0$ and $\dot{\theta}_0$ are initial conditions for the pendulum, and the terms inside the braces $\lbrace \cdot\cdot\cdot \rbrace$ describe decay for the overdamped case, and oscillations in the underdamped case.


If $\tau$ is fixed in time, for large damping $\zeta$, then the oscillator will converge to its new equilibrium position exponentially in time with a rate $\zeta \omega_0$ according to the solution \eqref{homosol}.  After this time, the angular displacement can be approximated by the fixed point of \eqref{theq}, given by
\be \label{theta-fp}
\bar{\theta} = \frac{\tau}{I \: \omega_0^2} = \frac{\tau}{\kappa}.
\ee
\par 
We wish to design the pendulum to respond sensitively to stimuli from the target objects, but do not want it to oscillate for a long time before returning to a new equilibrium position. There is a trade-off between sensitivity of the measurement and the speed of the response as will be explored in the following sections.

\subsection{Pendulum Model}
The pendulum is fixed with respect to its center of mass motion, and is allowed to only rotate about its center of mass in the plane of the figure.  We analyze this geometry by computing the torque about the middle of the torsion pendulum.  

Plane trigonometry dictates that the distances defined in Fig.~\ref{fig-geometry-pendulum} are given by 
\be
r_1^2 = (x- \ell)^2 + d^2, \quad r_2^2 = (x
+\ell)^2 + d^2, \quad r_0^2 = x^2 + d^2.
\ee
The gravitational force between mass $M$ and mass $m_j$, according to Newton \cite{newton1687}, is given by
\be
{\bf  F}_j = -\frac{ G m_j M}{r_j^2}{\bf \hat r}_j.
\ee
The torque $\tau$ generated on mass $j$ is given by 
\be
\tau_j =  \ell F_j \cos \theta_j,
\ee
where $F_j$ is the magnitude of gravitational force of mass $M$ on mass $m_j$.

For a Simple Pendulum (SP), we only have one of the masses in the pendulum of Fig.~\ref{fig-geometry-pendulum}.  The net torque is: 
\be
\tau_{SP} =   G m M \frac{ \ell d}{r_1^3}. \label{one-arm}
\ee
The above results are relevant for a single test mass other than the pendulum mass.  In the following sections we will use either the earth's gravity as the restoring force (as usual pendulums do), or by orienting the pendulum perpendicular to the earth's field, can also use the restoring force of a rod to obtain longer periods.

For a balanced Torsion Pendulum (TP), we include torques that nearly counterbalance each other (the torque on mass 1 is positive in sign, and the torque on mass 2 is negative in sign).  The net torque is given by
\be
\tau_\Sigma = \sum_j \tau_j = \tau_{TP} = \ell d G m M \left( \frac{1}{r_1^3} - \frac{1}{r_2^3}\right),  \label{torsion-torque}
\ee
where we have replaced $\cos \theta_j = d/r_j$.

The simple pendulum responds to the bare force on the sensing mass, and thus decays as $1/r^2$ with respect to the test mass distance.  The torsion pendulum balances the average force, and thus responds to the gradient of the field across the size of the torsion pendulum.  This effect leads to a less sensitive response to objects far away; it may be beneficial since it efficiently screens out far away objects and allows the sensor to focus on nearby objects.

\subsection{Limiting case}
In some experiments, we can further simplify the expression (\ref{torsion-torque}), since we expect that $\ell \ll d, x, r_0$ for some applications of interest.  The expression for $\tau_\Sigma$ is proportional to the difference of the functions $g(\ell) - g(-\ell)$, where 
\be
g(\ell) = \frac{1}{(d^2 + (x-\ell)^2)^{3/2}}.
\ee
Since $\ell$ is a small parameter, we can approximate $g(\ell) - g(-\ell) \approx g'(0) (2\ell)$.  We find that $g'(0) = 3 x/r_0^5$, so that we have to a good approximation,
\be
\tau_\Sigma = \frac{6 G m M d \ell^2  x}{r_0^5} = \frac{6 G m M \ell^2 \cos \theta \sin \theta}{r_0^3}, \label{far}
\ee
where we approximate $\theta_1 \approx \theta_2 = \theta$, and write $x = r_0 \sin \theta$ and $d = r_0 \cos \theta$.  In this limit, the sensor does not respond to the net force, but rather to its gradient, as indicated by the $r^{-3}$ law.

\par In this limit, the one--armed device (SP) equilibrates to an angle 
\be 
\bar{\theta}_{SP} \approx \frac{G M \cos\theta}{\ell \omega_0^2 r_0^2}, \label{1armdef}
\ee 
while its two--armed counterpart equilibrates to 
\be \label{2arm}
\bar{\theta}_\Sigma 
\approx \frac{3 G M \sin\theta \cos\theta}{\omega_0^2 r_0^3}.
\ee
In both cases we have used Eq.~\eqref{theta-fp}. We stress that in both cases, the sensing mass $m$ only appears in the natural frequency, and in the case of the torsion pendulum, the length $\ell$ also drops out, indicating that small sensors work as well as large ones so long as their periods are the same.

\begin{figure*}
\begin{picture}(450,200)
\put(0,0){\includegraphics[width = .48\textwidth]{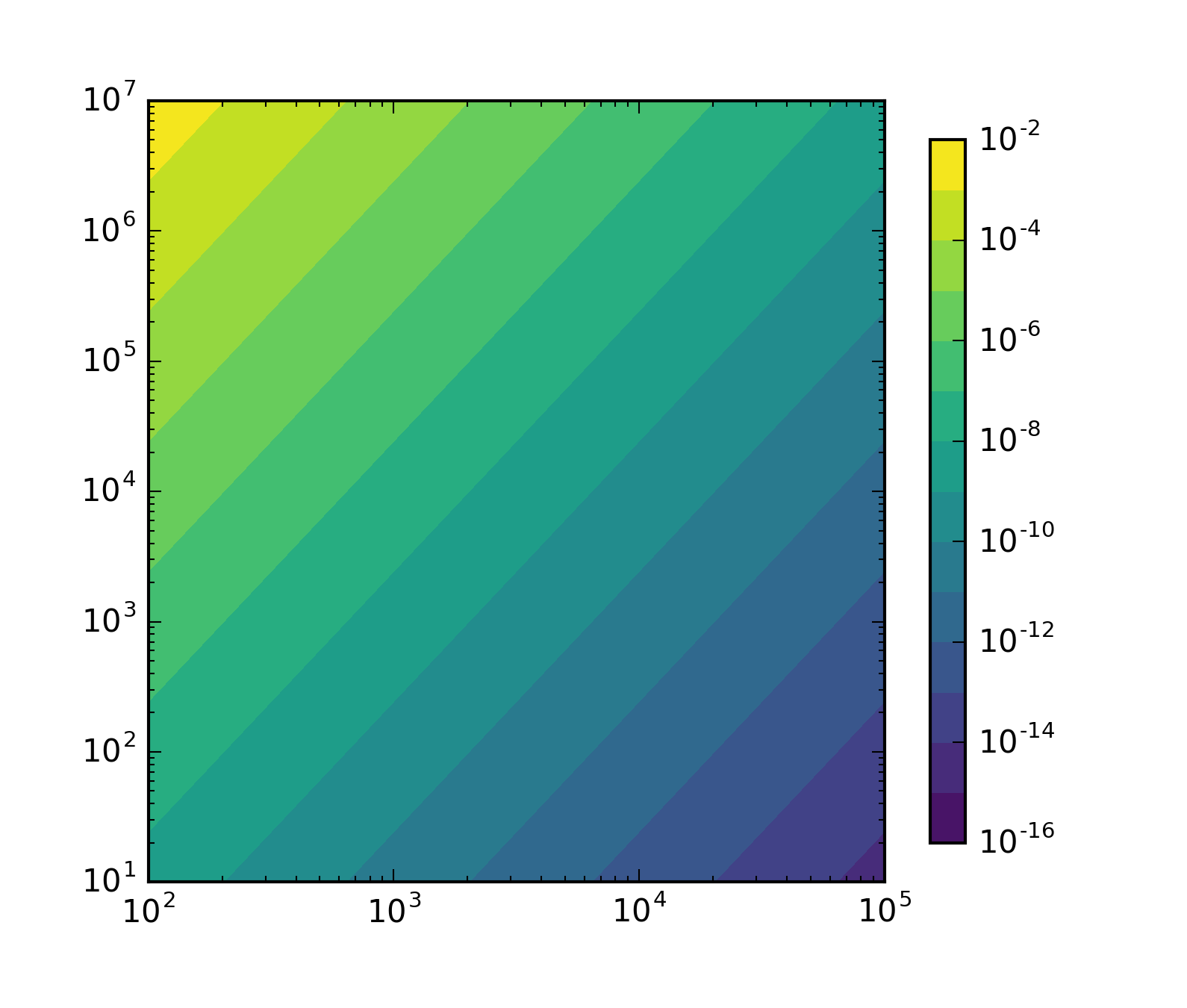}}
\put(232,0){\includegraphics[width = .48\textwidth]{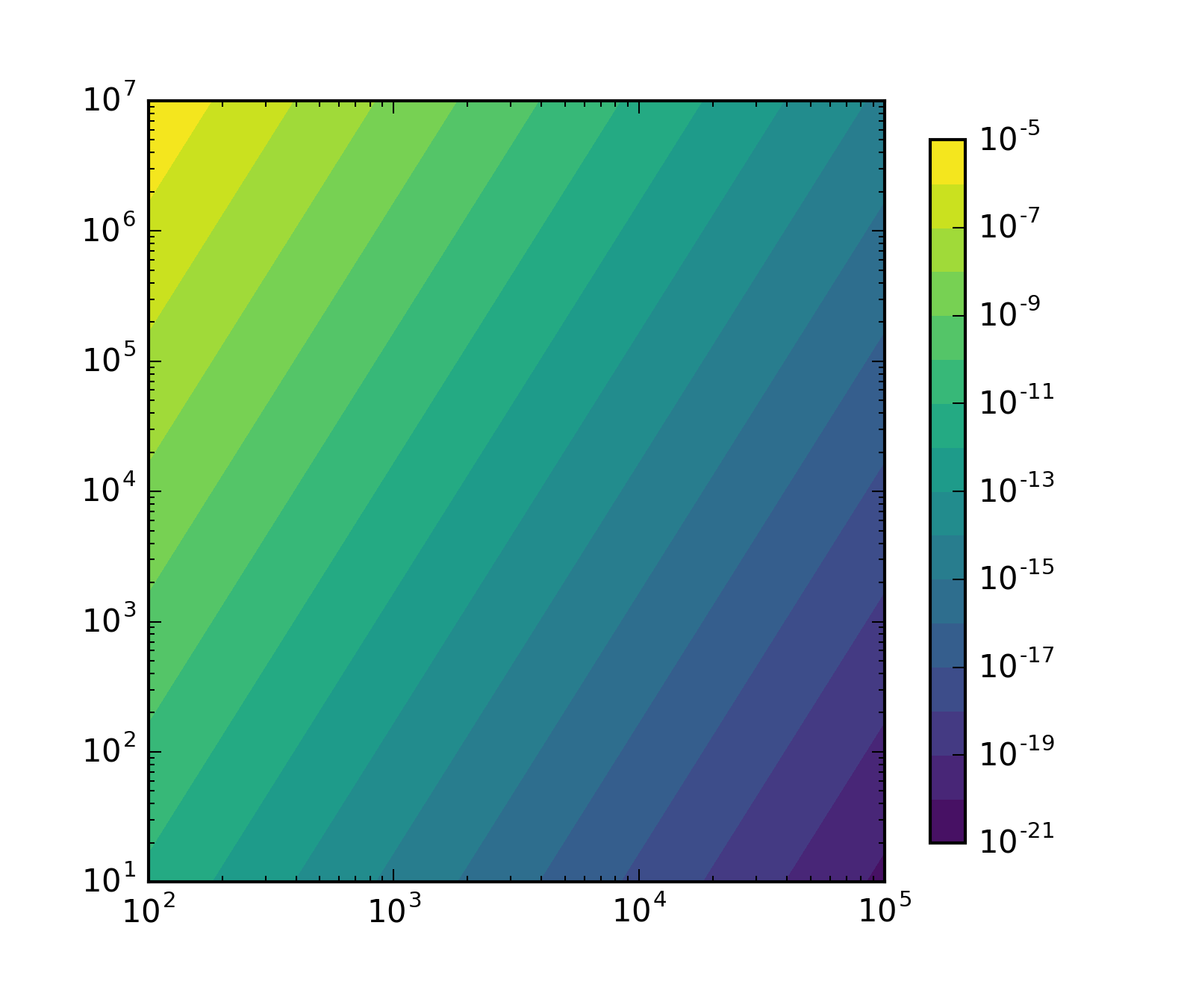}}
\put(-2,172){$M$ (kg)}
\put(230,172){$M$ (kg)}
\put(94,10){$r_0$ (m)}
\put(327,10){$r_0$ (m)}
\put(190,190){$\bar{\theta}_{SP}$ (rad)}
\put(420,190){$\bar{\theta}_\Sigma$ (rad)}
\end{picture} \vspace{-10pt}
\caption{We plot the static deflection angle $\bar{\theta}$ (\ref{theta-fp}) for the torque in the one-armed torsion pendulum \eqref{1armdef}, (left) and the two-armed torsion pendulum \eqref{2arm}, (right) as shown on the colorbars, as a function of target distance $r_0$ ($x$-axis) and target mass $M$ ($y$-axis). The plot, given as a log-log-log density plot, shows three decades of distance, and six of mass. The test mass $M$ is placed at the point of optimal sensitivity for each device.
The parameters in Table I are used for these plots. 
\label{fig-staticdef}}
\end{figure*}
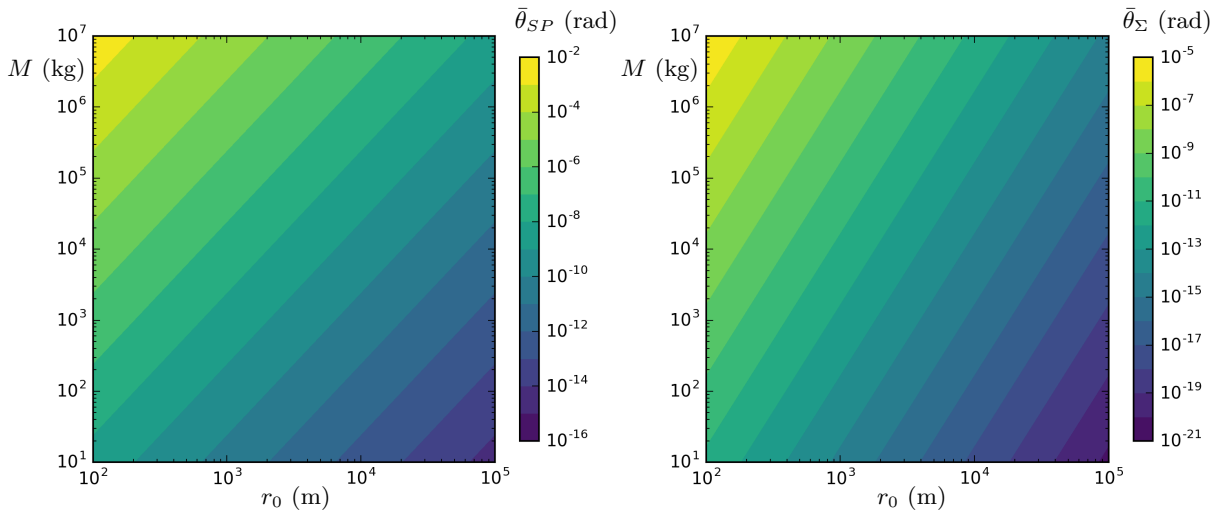

These expressions can be applied to make an approximate survey of the sensitivity of the device to different objects. Specifically, we show the best--case angular response to a target $M$ at distance $r_0$ in Fig.~\ref{fig-staticdef}, and we plot the angular dependence of the sensing for each device in Fig.~\ref{fig-angular}.
\begin{table}[ht]
\caption{Example parameter values} 
\centering 
\begin{tabular}{|c|c|c|}
\hline
Pendulum mass & $m$ & \unit[100]{g} \\
Wavelength of light & $\lambda$ & \unit[500]{nm} \\
Pendulum length & $\ell$ & \unit[5]{cm}\\
Period of oscillator & $\cal T$ & \unit[500]{s}\\
Torsion spring constant & $\kappa$ &  \unit[7.9 $\times 10^{-8}$]{kg m$^2$/s$^2$}\\
Length between beams on the mirror & L & \unit[1]{cm} \\
\hline
\end{tabular}
\label{table:values} 
\end{table}
Some example values for a small torsion pendulum are given in Table I, in reference to the geometry of Fig.~\ref{fig-geometry-pendulum}.  


The previous analysis may be extended to a continuous mass distribution by replacing the mass $M$ by a differential element $dM = \rho(x) dx$, where we imagine a body with mass per unit distance $\rho(x)$ distributed along the $x$ direction.  In that case, the next torque for such a mass distribution is given by
\be
\tau_\Sigma =  \int f(x) \rho(x) dx.
\ee
In the general case, $f(x) =\ell d G m (r_1(x)^{-3} - r_2(x)^{-3})$, whereas in the limiting case, it is given by $f =6 \ell^2 d G m x/(d^2 + x^2)^{5/2}$.

\section{Noise Considerations}\label{noise}

As for noise sources in the problem, we note that the pendulum will experience several kinds of noise that must be mitigated in order to reach the fundamental limits of angle detection that the system is capable of.  We focus on three type of noise in this section:  thermal noise, measurement heating noise, and quantum noise of the oscillator.

\subsection{Thermal noise}
Contributions of the thermal noise from the surrounding environment can be computed via the equipartition theorem assuming large temperatures.  Both the mean kinetic energy and potential energy are given by the thermal energy for one degree of freedom each.  In general,
\be
\frac{1}{2} \kappa \la \delta \theta^2 \ra = \frac{\hbar\omega}{4}\coth(\hbar\omega/2k_B T). \label{statmech}
\ee
In the limit of high temperatures, the equipartition of potential energy indicates that 
\be
\frac{1}{2} \kappa \la \delta \theta^2 \ra = \frac{1}{2} k_B T, 
\ee
which gives the typical rms noise of the torsion pendulum,
\be
\delta \theta_{rms} = \sqrt{k_B T/\kappa}.
\ee
  We can estimate the value using the values given in Table I, and room temperature $k_B T = \unit[4.1 \times 10^{-21}]{J}$ to find, $\delta \theta_{rms} = \unit[2.3 \times 10^{-7}]{rad}$.  In order to access below the picoradian regime, it will therefore be necessary to either cool the oscillator, or to time--average the signal for some time. One could also increase the value of $\kappa$, but that would also decrease the angular precision as well.

\begin{figure*}
\begin{tabular}{cc}
\includegraphics[width=.49\textwidth,trim = {10pt 15pt 20pt 7pt},clip]{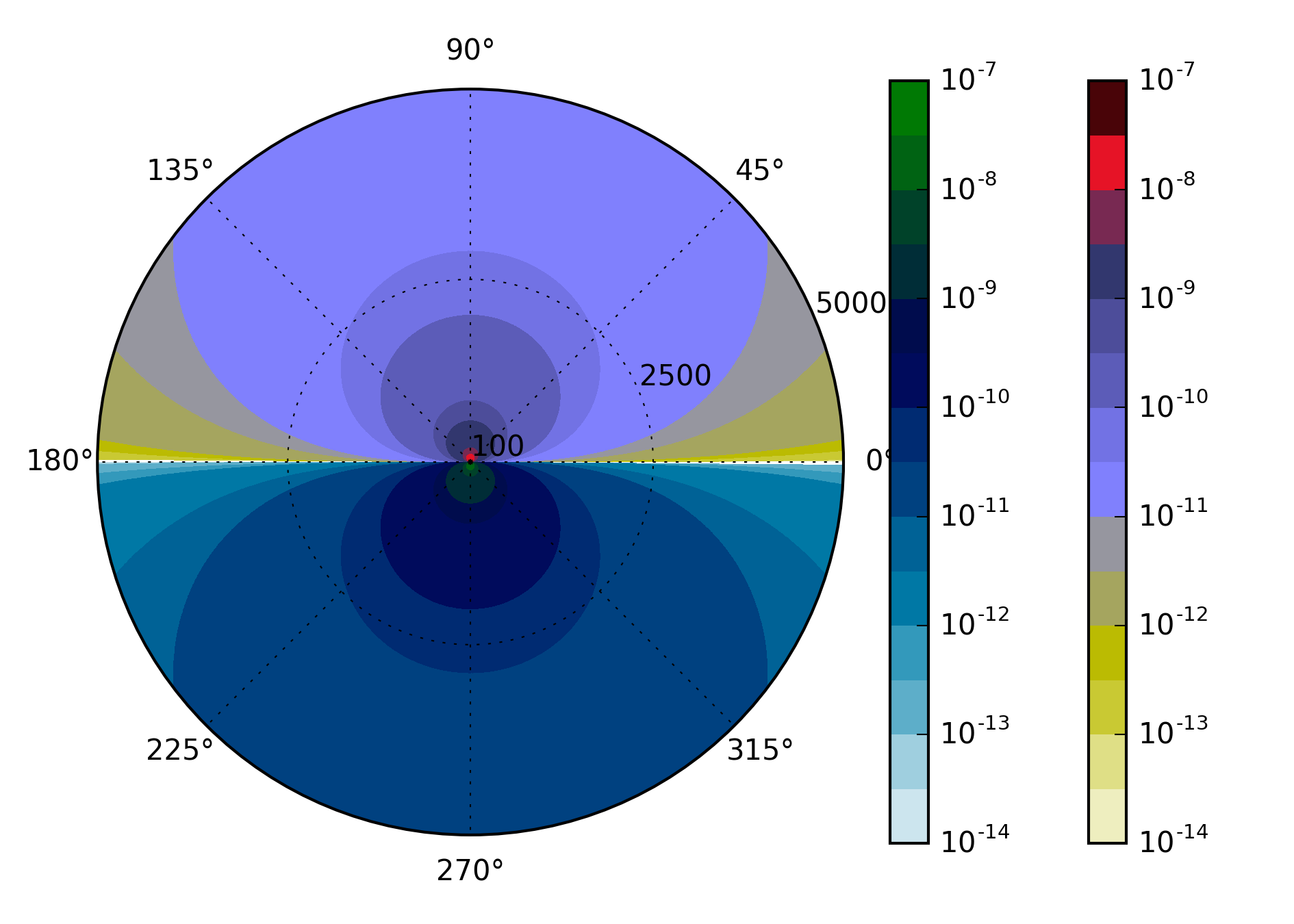} &
\includegraphics[width=.49\textwidth,trim = {10pt 15pt 20pt 7pt},clip]{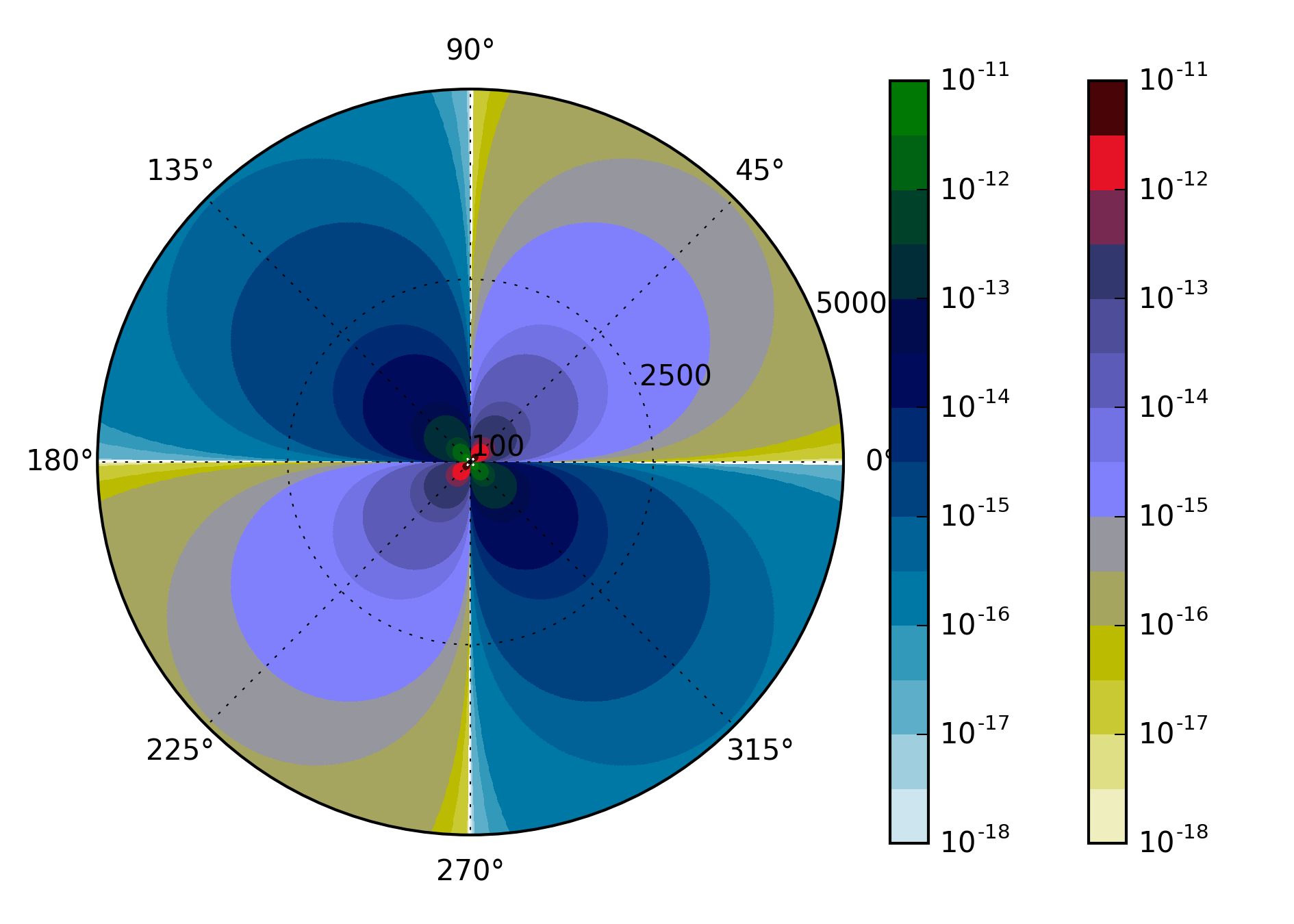}
\end{tabular} \begin{picture}(0,0)
\put(-32,173){$\boldsymbol{+}$}
\put(222,173){$\boldsymbol{+}$}
\put(182,173){$\boldsymbol{-}$}
\put(-72,173){$\boldsymbol{-}$}
\put(-68,183){$\bar{\theta}_{SP}$ (rad)}
\put(189,183){$\bar{\theta}_\Sigma$ (rad)}
\put(-120,10){1--arm}
\put(140,10){2--arm}
\end{picture}
\caption{
We show contour plots of the angular displacement $\bar{\theta}$ due to a static mass $M = 100~\mathrm{kg}$ placed in the plane of a 1--armed (left) and 2--armed pendulum (right), as a function of $r_0$ and $\theta$. We show the entire angular dependence $\theta$, and show values of $r_0$ ranging from $100~\mathrm{m}$ to $5~\mathrm{km}$ in the radial direction. The $+$ colorbar denotes an angular displacement in the $+\theta$ (CCW) direction, while the $-$ colorbar denotes deflection in the $-\theta$ (CW) direction. Angular blindspots are at the juncture of the two colorbars, where the deflection is zero, no matter the value of $r_0$ or $M$. Numerical values for the pendulum correspond to those shown in Table \ref{table:values}. 
}\label{fig-angular}
\end{figure*}
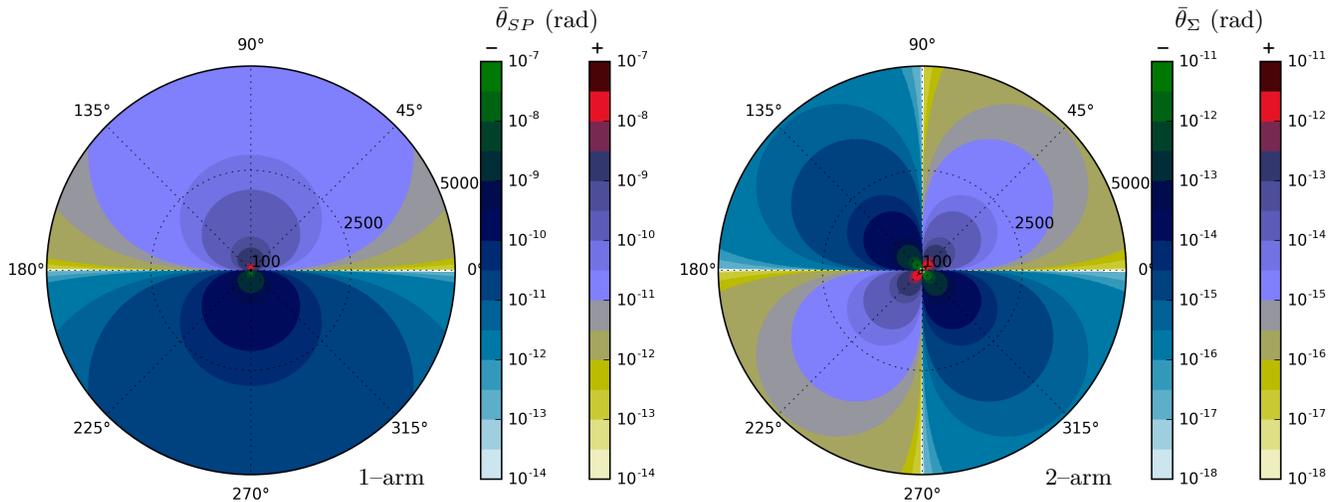

\subsection{Measurement heating}
As demonstrated in the first section, it is advantageous to apply as much optical power as possible to the interferometer to maximize the precision of the angle measurement.
However, because the gravitational sensor is freely moving, it is possible that the sensing laser may drive excitations, effectively heating the torsion pendulum.  We will now calculate the effect of this heat, which may put a bound on the sensing power.

\par The displaced Sagnac geometry (see Fig.~\ref{model}), causes two laser beams to strike the sensing mirror at a lever arm $\pm\frac{L}{2}$ from the axis of rotation. We can imagine that photons each carrying energy $h c/ \lambda = \hbar k_0 c$ arrive at the mirror at discrete times, either at $+\tfrac{L}{2}$ or $-\tfrac{L}{2}$. 
On average, since the intensity of the light on the left and right side of the mirror is the same from the 50/50 beamsplitter, we have $\la N_+ \ra = \la N_- \ra = \tfrac{N}{2}$. 
The time--averaged optical power being applied to the pendulum is $\bar{P} = \la N \ra \hbar k_0 c / T$, which could be understood as arising from
\be 
\bar{P}_\pm  = \frac{1}{T}\int_0^T \hbar k_0 c \left( \sum_{n}^{N_\pm} \delta(t-t_n) \right) dt,
\ee
provided that the duration $T$ is much longer than the typical time between individual photons, which arrive at the $t_n$, and where $\bar{P} = \bar{P}_+ + \bar{P}_-$. 
Each photon which strikes the mirror imparts an impulse $\Delta p = \hbar k_0/\sqrt{2}$, where the factor $\sqrt{2}$ arises because the photons strike the mirror at an angle. 
The time--averaged torque may be expressed using the same logic as above, i.e.
\be 
\bar{\tau}_\pm = \frac{1}{T} \int_0^T \frac{L}{2} \Delta p \left( \sum_{n}^{N_\pm} \delta(t-t_n) \right) dt,
\ee
or for $T \gg |t_n - t_{n-1}|$,
\be 
\la \bar{\tau} \ra = \la \bar{\tau}_+ - \bar{\tau}_- \ra  = \frac{L k_0 \hbar}{T \sqrt{2}} \la N_+ - N_- \ra. 
\ee
Since the number of photons striking on either side of the point of rotation is equal on average, we see that $\la \bar{\tau} \ra = 0$;
however, there will be {\em fluctuations} from the coherent states of light, which depend on
\be 
\left\langle (N_+ - N_-)^2 \right\rangle = \left\langle N_+^2 + N_-^2 - 2 N_+ N_- \right\rangle = \la N \ra. 
\ee
The term $\la N_+ N_- \ra = 0$ because the $\pm$ photon beams are uncorrelated. 
Furthermore, given the geometry and using statistical properties of coherent states, $\la N_+^2\ra=\la N_+\ra=\frac{N}{2}$
and $\la N_-^2\ra=\la N_-\ra=\frac{N}{2}$. Therefore the torque fluctuations averaged over a given time interval $T$ are given by
\be \begin{split}
\la \bar{\tau}^2 \ra &= \frac{L^2 k_0^2 \hbar^2}{2 T^2}\left\langle (N_+ - N_-)^2 \right\rangle =  \frac{L^2 k_0^2 \hbar^2}{2 T^2} \la N \ra 
 = \frac{L^2 \bar{P}^2}{2 c^2 \la N \ra},
\end{split} \ee
where we have re--written the duration of the experiment in terms of the power $\bar{P}$, and used the relationships above.
This leads to an increased variance of the angle, defined by $\la\delta \bar{\theta}^2\ra = \la \bar{\tau}^2\ra/\kappa^2$, or taking the square root, fluctuations in the coherent states of light lead to
\be 
\delta \bar{\theta} =  \frac{L \bar{P}}{\kappa c \sqrt{2 N}}.
\ee
Equating with the shot noise limited value \eqref{res-limit}, we find that the measurement heating makes a smaller contribution to the noise than shot noise, when the condition
\be 
\frac{L^2 \bar{P} k_0}{\kappa} \leq \frac{c}{2}
\ee
is satisfied. This implies that for the values in Table~\ref{table:values}, the noise contribution due to measurement heating would overtake the shot noise when the measurement laser power exceeds $\bar{P} = \unit[9.3]{mW}$.

\subsection{Quantum Noise}
An intriguing aspect of the oscillators is the fundamental limitation of sensitivity due to quantum noise. As will be shown, the shot noise limited resolution is approximately equal to the quantum ground state uncertainty of the oscillator when the integration time is approximately equal to the period of the pendulum.  From a quantum mechanical perspective, ground state quantum noise limitations are quite interesting in light of the large masses used in these experiment.  Such studies may be valuable in probing quantum gravity.  On the other hand, this also places fundamental noise limits on the resolution.

To determine the ground state angular uncertainty, we set the mean potential energy of the oscillator to the ground state energy of the oscillator
\begin{equation}
(1/2) \kappa \la (\delta \theta)^2 \ra = (1/4)  \hbar \omega,
\end{equation}
which follows from Eq.~\eqref{statmech} when $k_b T \ll \hbar \omega$.
Solving for the angle we obtain
\be
\delta \theta_{rms} = \sqrt{\hbar \omega/2\kappa}.
\ee
Using values listed in Table \ref{table:values}, we find $\delta \theta_{rms}=\unit[2.9 \times 10^{-15}]{rad}$. This resolution can be achieved when the integration time is roughly equal to the period of the pendulum using a few milliwatts of laser power.

\section{Limits of resolution}
\label{resolution}
The results of the previous sections can now be combined to give the sensitivity limits of the SP and TP to forces, which can be translated into either mass or range uncertainty. Using Eq.~(\ref{theta-fp}) and the angular uncertainty Eq.~\ref{res-limit}, we find at the optimally sensitive response point ($x=0$ so $d \approx r$, $\theta=0$), the resolution of (usual) simple pendulum acceleration $a$, relative to the gravitational acceleration $g$ of the SP to be
\be
\frac{\delta a}{g} = \delta \theta.
\ee
Consequently, the acceleration uncertainty in units of the accelerations due to gravity near the surface of the earth is simply the same as the angular uncertainty.

If instead, we consider a one-armed torsion pendulum with a torsion constant of $\kappa$, oriented perpendicular to the gravitational field of the earth, then the period of the oscillation can be much longer.  The angular resolution is given by Eq.~\eqref{1armdef} so the acceleration uncertainty is reduced to
\be
\frac{\delta a}{g}  = \frac{\kappa \delta \theta}{g \ell m}.
\ee
For the parameters in Table I, this reduces the acceleration uncertainty by a factor of $1.6 \times 10^{-6}$, leading to \unit[60]{zepto g Hz$^{-1/2}$}.  Remarkably, the speed is only a thousand times slower, because of the inverse square relationship of Eq.~\eqref{kappa}.   In either geometry, the acceleration is given by $a = G M /r_0^2$, so the sensitivity of the acceleration to a change in test mass $\delta M$ at fixed $r_0$, or a change in the distance $\delta r_0$ for fixed test mass $M$ is given by
\be
\delta a = G \delta M/r_0^2  - 2 G M \delta r_0/r_0^3,
\ee
from which the mass or distance uncertainty is easily found.  The response of the one-armed torsion pendulum is plotted in Fig.~\ref{fig-staticdef}(left) for different values of test mass $M$ and range $R=r_0$.

For a balanced torsion pendulum, a test mass far from the pendulum will respond according to Eqs.~(\ref{far},\ref{2arm}).  Setting $\theta=\pi/4$ for maximum sensitivity, the angular response to a gravitating body will be
\be
\delta \theta  =  \frac{3 G m M \ell^2}{\kappa r_0^3}.
\ee
The $r_0^{-3}$ law gives a smaller sensitivity, but also screens off distant objects.  This cannot be directly translated into acceleration of a single mass, but gives the response of the detector to the gradient of the gravitational field.  The torsion pendulum response is plotted in Fig.~\ref{fig-staticdef}(right) for different values of test mass $M$ and range $R=r_0$.

We now briefly discuss the angular response of both types of pendula as the test mass is placed at different angles relative to the axis of rotation.  The one-armed torsion pendulum has blind spots at $\theta \approx 0$ and $\theta \approx \pi$, where a target mass applies no torque, and its sensitivity is maximized at $\theta \approx \pi/2$ and $\theta \approx 3\pi/2$. The two--armed torsion pendulum has four blind--spots, as illustrated in Fig.~\ref{fig-angular}. Notice that the scaling of the deflection angle in terms of the one-armed pendulum's construction parameters really depends only on $\ell$, and that the the smaller we make $\ell$, the larger the deflection angle will get (the moment of inertia in the denominator wins out over the greater torque with greater arm length). 
The $\ell$ dependence cancels out entirely from the two--armed device, except for its appearance in the natural frequency.

\section{Conclusions} \label{conc}
We have shown how a sensitive gravitational sensor can be built using advanced optical interferometry techniques.  By allowing a mechanical element to oscillate freely and including a mirror on this element, which is incorporated into the interferometer, a slight tilt of the mirror causes the counter--propagating optical beams in the interferometer to acquire a phase difference between each other.  That phase difference can then be read out with an inverse weak value technique.  This method results in a double--lobe distribution whose mean sensitively depends on the phase, which in turn depends on the angular tilt of the mirror.  Our analysis indicates that we can reach acceleration sensitivities of tens of zepto-g per root-Hertz for \unit[1]{mW} of power.  We have discussed how that sensing threshold can be traded between mass and range of targets.

\section{Acknowledgements}
ANJ and PL acknowledge funding from Leonardo DRS technologies, and a University of Rochester pump-primer award. JCH acknowledges funding from ARO. JT acknowledges support by the Fetzer Franklin Fund of the John E.~Fetzer Memorial Trust. ANJ discloses that a portion of this research was conducted outside of the University of Rochester through his LLC. Financial interests include ownership and fiduciary roles in the LLC. PL and ANJ would like to thank Kevin Lyons for helpful discussions.
We thank the Institute for Quantum Studies at Chapman University for support. We also thank Steven and Jennifer Baker of Laguna Beach for their hospitality during the writing of this manuscript.

\bibliography{references}

\begin{thebibliography}{47}%
\makeatletter
\providecommand \@ifxundefined [1]{%
 \@ifx{#1\undefined}
}%
\providecommand \@ifnum [1]{%
 \ifnum #1\expandafter \@firstoftwo
 \else \expandafter \@secondoftwo
 \fi
}%
\providecommand \@ifx [1]{%
 \ifx #1\expandafter \@firstoftwo
 \else \expandafter \@secondoftwo
 \fi
}%
\providecommand \natexlab [1]{#1}%
\providecommand \enquote  [1]{``#1''}%
\providecommand \bibnamefont  [1]{#1}%
\providecommand \bibfnamefont [1]{#1}%
\providecommand \citenamefont [1]{#1}%
\providecommand \href@noop [0]{\@secondoftwo}%
\providecommand \href [0]{\begingroup \@sanitize@url \@href}%
\providecommand \@href[1]{\@@startlink{#1}\@@href}%
\providecommand \@@href[1]{\endgroup#1\@@endlink}%
\providecommand \@sanitize@url [0]{\catcode `\\12\catcode `\$12\catcode
  `\&12\catcode `\#12\catcode `\^12\catcode `\_12\catcode `\%12\relax}%
\providecommand \@@startlink[1]{}%
\providecommand \@@endlink[0]{}%
\providecommand \url  [0]{\begingroup\@sanitize@url \@url }%
\providecommand \@url [1]{\endgroup\@href {#1}{\urlprefix }}%
\providecommand \urlprefix  [0]{URL }%
\providecommand \Eprint [0]{\href }%
\providecommand \doibase [0]{http://dx.doi.org/}%
\providecommand \selectlanguage [0]{\@gobble}%
\providecommand \bibinfo  [0]{\@secondoftwo}%
\providecommand \bibfield  [0]{\@secondoftwo}%
\providecommand \translation [1]{[#1]}%
\providecommand \BibitemOpen [0]{}%
\providecommand \bibitemStop [0]{}%
\providecommand \bibitemNoStop [0]{.\EOS\space}%
\providecommand \EOS [0]{\spacefactor3000\relax}%
\providecommand \BibitemShut  [1]{\csname bibitem#1\endcsname}%
\let\auto@bib@innerbib\@empty
\bibitem [{\citenamefont {Aharonov}\ \emph {et~al.}(1988)\citenamefont
  {Aharonov}, \citenamefont {Albert},\ and\ \citenamefont
  {Vaidman}}]{Aharonov88}%
  \BibitemOpen
  \bibfield  {author} {\bibinfo {author} {\bibfnamefont {Yakir}\ \bibnamefont
  {Aharonov}}, \bibinfo {author} {\bibfnamefont {David~Z.}\ \bibnamefont
  {Albert}}, \ and\ \bibinfo {author} {\bibfnamefont {Lev}\ \bibnamefont
  {Vaidman}},\ }\bibfield  {title} {\enquote {\bibinfo {title} {How the result
  of a measurement of a component of the spin of a spin-1/2 particle can turn
  out to be 100},}\ }\href {\doibase 10.1103/PhysRevLett.60.1351} {\bibfield
  {journal} {\bibinfo  {journal} {Phys. Rev. Lett.}\ }\textbf {\bibinfo
  {volume} {60}},\ \bibinfo {pages} {1351--1354} (\bibinfo {year}
  {1988})}\BibitemShut {NoStop}%
\bibitem [{\citenamefont {Ritchie}\ \emph {et~al.}(1991)\citenamefont
  {Ritchie}, \citenamefont {Story},\ and\ \citenamefont {Hulet}}]{Ritchie91}%
  \BibitemOpen
  \bibfield  {author} {\bibinfo {author} {\bibfnamefont {N.~W.~M.}\
  \bibnamefont {Ritchie}}, \bibinfo {author} {\bibfnamefont {J.~G.}\
  \bibnamefont {Story}}, \ and\ \bibinfo {author} {\bibfnamefont {Randall~G.}\
  \bibnamefont {Hulet}},\ }\bibfield  {title} {\enquote {\bibinfo {title}
  {{Realization of a measurement of a `weak value'}},}\ }\href {\doibase
  10.1103/PhysRevLett.66.1107} {\bibfield  {journal} {\bibinfo  {journal}
  {Phys. Rev. Lett.}\ }\textbf {\bibinfo {volume} {66}},\ \bibinfo {pages}
  {1107--1110} (\bibinfo {year} {1991})}\BibitemShut {NoStop}%
\bibitem [{\citenamefont {Hosten}\ and\ \citenamefont
  {Kwiat}(2008)}]{Hosten08}%
  \BibitemOpen
  \bibfield  {author} {\bibinfo {author} {\bibfnamefont {Onur}\ \bibnamefont
  {Hosten}}\ and\ \bibinfo {author} {\bibfnamefont {Paul}\ \bibnamefont
  {Kwiat}},\ }\bibfield  {title} {\enquote {\bibinfo {title} {Observation of
  the spin hall effect of light via weak measurements},}\ }\href {\doibase
  10.1126/science.1152697} {\bibfield  {journal} {\bibinfo  {journal}
  {Science}\ }\textbf {\bibinfo {volume} {319}},\ \bibinfo {pages} {787--790}
  (\bibinfo {year} {2008})}\BibitemShut {NoStop}%
\bibitem [{\citenamefont {Dixon}\ \emph {et~al.}(2009)\citenamefont {Dixon},
  \citenamefont {Starling}, \citenamefont {Jordan},\ and\ \citenamefont
  {Howell}}]{Dixon09}%
  \BibitemOpen
  \bibfield  {author} {\bibinfo {author} {\bibfnamefont {P.~Ben}\ \bibnamefont
  {Dixon}}, \bibinfo {author} {\bibfnamefont {David~J.}\ \bibnamefont
  {Starling}}, \bibinfo {author} {\bibfnamefont {Andrew~N.}\ \bibnamefont
  {Jordan}}, \ and\ \bibinfo {author} {\bibfnamefont {John~C.}\ \bibnamefont
  {Howell}},\ }\bibfield  {title} {\enquote {\bibinfo {title} {Ultrasensitive
  beam deflection measurement via interferometric weak value amplification},}\
  }\href {\doibase 10.1103/PhysRevLett.102.173601} {\bibfield  {journal}
  {\bibinfo  {journal} {Phys. Rev. Lett.}\ }\textbf {\bibinfo {volume} {102}},\
  \bibinfo {pages} {173601} (\bibinfo {year} {2009})}\BibitemShut {NoStop}%
\bibitem [{\citenamefont {Brunner}\ and\ \citenamefont
  {Simon}(2010)}]{Brunner10}%
  \BibitemOpen
  \bibfield  {author} {\bibinfo {author} {\bibfnamefont {Nicolas}\ \bibnamefont
  {Brunner}}\ and\ \bibinfo {author} {\bibfnamefont {Christoph}\ \bibnamefont
  {Simon}},\ }\bibfield  {title} {\enquote {\bibinfo {title} {Measuring small
  longitudinal phase shifts: Weak measurements or standard interferometry?}}\
  }\href {\doibase 10.1103/PhysRevLett.105.010405} {\bibfield  {journal}
  {\bibinfo  {journal} {Phys. Rev. Lett.}\ }\textbf {\bibinfo {volume} {105}},\
  \bibinfo {pages} {010405} (\bibinfo {year} {2010})}\BibitemShut {NoStop}%
\bibitem [{\citenamefont {Dressel}\ \emph {et~al.}(2013)\citenamefont
  {Dressel}, \citenamefont {Lyons}, \citenamefont {Jordan}, \citenamefont
  {Graham},\ and\ \citenamefont {Kwiat}}]{Dressel2013}%
  \BibitemOpen
  \bibfield  {author} {\bibinfo {author} {\bibfnamefont {Justin}\ \bibnamefont
  {Dressel}}, \bibinfo {author} {\bibfnamefont {Kevin}\ \bibnamefont {Lyons}},
  \bibinfo {author} {\bibfnamefont {Andrew~N.}\ \bibnamefont {Jordan}},
  \bibinfo {author} {\bibfnamefont {Trent~M.}\ \bibnamefont {Graham}}, \ and\
  \bibinfo {author} {\bibfnamefont {Paul~G.}\ \bibnamefont {Kwiat}},\
  }\bibfield  {title} {\enquote {\bibinfo {title} {Strengthening weak-value
  amplification with recycled photons},}\ }\href {\doibase
  10.1103/PhysRevA.88.023821} {\bibfield  {journal} {\bibinfo  {journal} {Phys.
  Rev. A}\ }\textbf {\bibinfo {volume} {88}},\ \bibinfo {pages} {023821}
  (\bibinfo {year} {2013})}\BibitemShut {NoStop}%
\bibitem [{\citenamefont {Lyons}\ \emph {et~al.}(2015)\citenamefont {Lyons},
  \citenamefont {Dressel}, \citenamefont {Jordan}, \citenamefont {Howell},\
  and\ \citenamefont {Kwiat}}]{Lyons2015}%
  \BibitemOpen
  \bibfield  {author} {\bibinfo {author} {\bibfnamefont {Kevin}\ \bibnamefont
  {Lyons}}, \bibinfo {author} {\bibfnamefont {Justin}\ \bibnamefont {Dressel}},
  \bibinfo {author} {\bibfnamefont {Andrew~N.}\ \bibnamefont {Jordan}},
  \bibinfo {author} {\bibfnamefont {John~C.}\ \bibnamefont {Howell}}, \ and\
  \bibinfo {author} {\bibfnamefont {Paul~G.}\ \bibnamefont {Kwiat}},\
  }\bibfield  {title} {\enquote {\bibinfo {title} {Power-recycled
  weak-value-based metrology},}\ }\href {\doibase
  10.1103/PhysRevLett.114.170801} {\bibfield  {journal} {\bibinfo  {journal}
  {Phys. Rev. Lett.}\ }\textbf {\bibinfo {volume} {114}},\ \bibinfo {pages}
  {170801} (\bibinfo {year} {2015})}\BibitemShut {NoStop}%
\bibitem [{\citenamefont {Mart\'{i}nez-Rinc\'{o}n}\ \emph
  {et~al.}(2017)\citenamefont {Mart\'{i}nez-Rinc\'{o}n}, \citenamefont
  {Mullarkey}, \citenamefont {Viza}, \citenamefont {Liu},\ and\ \citenamefont
  {Howell}}]{Martinez17}%
  \BibitemOpen
  \bibfield  {author} {\bibinfo {author} {\bibfnamefont {Juli\'{a}n}\
  \bibnamefont {Mart\'{i}nez-Rinc\'{o}n}}, \bibinfo {author} {\bibfnamefont
  {Christopher~A.}\ \bibnamefont {Mullarkey}}, \bibinfo {author} {\bibfnamefont
  {Gerardo~I.}\ \bibnamefont {Viza}}, \bibinfo {author} {\bibfnamefont
  {Wei-Tao}\ \bibnamefont {Liu}}, \ and\ \bibinfo {author} {\bibfnamefont
  {John~C.}\ \bibnamefont {Howell}},\ }\bibfield  {title} {\enquote {\bibinfo
  {title} {Ultrasensitive inverse weak-value tilt meter},}\ }\href {\doibase
  10.1364/OL.42.002479} {\bibfield  {journal} {\bibinfo  {journal} {Opt.
  Lett.}\ }\textbf {\bibinfo {volume} {42}},\ \bibinfo {pages} {2479--2482}
  (\bibinfo {year} {2017})}\BibitemShut {NoStop}%
\bibitem [{\citenamefont {Starling}\ \emph {et~al.}(2009)\citenamefont
  {Starling}, \citenamefont {Dixon}, \citenamefont {Jordan},\ and\
  \citenamefont {Howell}}]{Starling09}%
  \BibitemOpen
  \bibfield  {author} {\bibinfo {author} {\bibfnamefont {David~J.}\
  \bibnamefont {Starling}}, \bibinfo {author} {\bibfnamefont {P.~Ben}\
  \bibnamefont {Dixon}}, \bibinfo {author} {\bibfnamefont {Andrew~N.}\
  \bibnamefont {Jordan}}, \ and\ \bibinfo {author} {\bibfnamefont {John~C.}\
  \bibnamefont {Howell}},\ }\bibfield  {title} {\enquote {\bibinfo {title}
  {Optimizing the signal-to-noise ratio of a beam-deflection measurement with
  interferometric weak values},}\ }\href {\doibase 10.1103/PhysRevA.80.041803}
  {\bibfield  {journal} {\bibinfo  {journal} {Phys. Rev. A}\ }\textbf {\bibinfo
  {volume} {80}},\ \bibinfo {pages} {041803} (\bibinfo {year}
  {2009})}\BibitemShut {NoStop}%
\bibitem [{\citenamefont {Feizpour}\ \emph {et~al.}(2011)\citenamefont
  {Feizpour}, \citenamefont {Xing},\ and\ \citenamefont
  {Steinberg}}]{Feizpour11}%
  \BibitemOpen
  \bibfield  {author} {\bibinfo {author} {\bibfnamefont {Amir}\ \bibnamefont
  {Feizpour}}, \bibinfo {author} {\bibfnamefont {Xingxing}\ \bibnamefont
  {Xing}}, \ and\ \bibinfo {author} {\bibfnamefont {Aephraim~M.}\ \bibnamefont
  {Steinberg}},\ }\bibfield  {title} {\enquote {\bibinfo {title} {Amplifying
  single-photon nonlinearity using weak measurements},}\ }\href {\doibase
  10.1103/PhysRevLett.107.133603} {\bibfield  {journal} {\bibinfo  {journal}
  {Phys. Rev. Lett.}\ }\textbf {\bibinfo {volume} {107}},\ \bibinfo {pages}
  {133603} (\bibinfo {year} {2011})}\BibitemShut {NoStop}%
\bibitem [{\citenamefont {Jordan}\ \emph {et~al.}(2014)\citenamefont {Jordan},
  \citenamefont {Mart\'{\i}nez-Rinc\'on},\ and\ \citenamefont
  {Howell}}]{Jordan14}%
  \BibitemOpen
  \bibfield  {author} {\bibinfo {author} {\bibfnamefont {Andrew~N.}\
  \bibnamefont {Jordan}}, \bibinfo {author} {\bibfnamefont {Juli\'an}\
  \bibnamefont {Mart\'{\i}nez-Rinc\'on}}, \ and\ \bibinfo {author}
  {\bibfnamefont {John~C.}\ \bibnamefont {Howell}},\ }\bibfield  {title}
  {\enquote {\bibinfo {title} {Technical advantages for weak-value
  amplification: When less is more},}\ }\href {\doibase
  10.1103/PhysRevX.4.011031} {\bibfield  {journal} {\bibinfo  {journal} {Phys.
  Rev. X}\ }\textbf {\bibinfo {volume} {4}},\ \bibinfo {pages} {011031}
  (\bibinfo {year} {2014})}\BibitemShut {NoStop}%
\bibitem [{\citenamefont {Viza}\ \emph {et~al.}(2015)\citenamefont {Viza},
  \citenamefont {Mart\'{\i}nez-Rinc\'on}, \citenamefont {Alves}, \citenamefont
  {Jordan},\ and\ \citenamefont {Howell}}]{Viza15}%
  \BibitemOpen
  \bibfield  {author} {\bibinfo {author} {\bibfnamefont {Gerardo~I.}\
  \bibnamefont {Viza}}, \bibinfo {author} {\bibfnamefont {Juli\'an}\
  \bibnamefont {Mart\'{\i}nez-Rinc\'on}}, \bibinfo {author} {\bibfnamefont
  {Gabriel~B.}\ \bibnamefont {Alves}}, \bibinfo {author} {\bibfnamefont
  {Andrew~N.}\ \bibnamefont {Jordan}}, \ and\ \bibinfo {author} {\bibfnamefont
  {John~C.}\ \bibnamefont {Howell}},\ }\bibfield  {title} {\enquote {\bibinfo
  {title} {Experimentally quantifying the advantages of weak-value-based
  metrology},}\ }\href {\doibase 10.1103/PhysRevA.92.032127} {\bibfield
  {journal} {\bibinfo  {journal} {Phys. Rev. A}\ }\textbf {\bibinfo {volume}
  {92}},\ \bibinfo {pages} {032127} (\bibinfo {year} {2015})}\BibitemShut
  {NoStop}%
\bibitem [{\citenamefont {Pang}\ \emph {et~al.}(2016)\citenamefont {Pang},
  \citenamefont {Alonso}, \citenamefont {Brun},\ and\ \citenamefont
  {Jordan}}]{pang2016protecting}%
  \BibitemOpen
  \bibfield  {author} {\bibinfo {author} {\bibfnamefont {Shengshi}\
  \bibnamefont {Pang}}, \bibinfo {author} {\bibfnamefont {Jose Raul~Gonzalez}\
  \bibnamefont {Alonso}}, \bibinfo {author} {\bibfnamefont {Todd~A.}\
  \bibnamefont {Brun}}, \ and\ \bibinfo {author} {\bibfnamefont {Andrew~N.}\
  \bibnamefont {Jordan}},\ }\bibfield  {title} {\enquote {\bibinfo {title}
  {Protecting weak measurements against systematic errors},}\ }\href {\doibase
  10.1103/PhysRevA.94.012329} {\bibfield  {journal} {\bibinfo  {journal} {Phys.
  Rev. A}\ }\textbf {\bibinfo {volume} {94}},\ \bibinfo {pages} {012329}
  (\bibinfo {year} {2016})}\BibitemShut {NoStop}%
\bibitem [{\citenamefont {Sinclair}\ \emph {et~al.}(2017)\citenamefont
  {Sinclair}, \citenamefont {Hallaji}, \citenamefont {Steinberg}, \citenamefont
  {Tollaksen},\ and\ \citenamefont {Jordan}}]{sinclair2017weak}%
  \BibitemOpen
  \bibfield  {author} {\bibinfo {author} {\bibfnamefont {Josiah}\ \bibnamefont
  {Sinclair}}, \bibinfo {author} {\bibfnamefont {Matin}\ \bibnamefont
  {Hallaji}}, \bibinfo {author} {\bibfnamefont {Aephraim~M.}\ \bibnamefont
  {Steinberg}}, \bibinfo {author} {\bibfnamefont {Jeff}\ \bibnamefont
  {Tollaksen}}, \ and\ \bibinfo {author} {\bibfnamefont {Andrew~N.}\
  \bibnamefont {Jordan}},\ }\bibfield  {title} {\enquote {\bibinfo {title}
  {Weak-value amplification and optimal parameter estimation in the presence of
  correlated noise},}\ }\href {\doibase 10.1103/PhysRevA.96.052128} {\bibfield
  {journal} {\bibinfo  {journal} {Phys. Rev. A}\ }\textbf {\bibinfo {volume}
  {96}},\ \bibinfo {pages} {052128} (\bibinfo {year} {2017})}\BibitemShut
  {NoStop}%
\bibitem [{\citenamefont {Lyons}\ \emph {et~al.}(2017)\citenamefont {Lyons},
  \citenamefont {Howell},\ and\ \citenamefont {Jordan}}]{Lyons2017}%
  \BibitemOpen
  \bibfield  {author} {\bibinfo {author} {\bibfnamefont {Kevin}\ \bibnamefont
  {Lyons}}, \bibinfo {author} {\bibfnamefont {John~C.}\ \bibnamefont {Howell}},
  \ and\ \bibinfo {author} {\bibfnamefont {Andrew~N.}\ \bibnamefont {Jordan}},\
  }\bibfield  {title} {\enquote {\bibinfo {title} {Noise suppression in inverse
  weak value-based phase detection},}\ }\href
  {https://doi.org/10.1007/s40509-017-0145-7} {\bibfield  {journal} {\bibinfo
  {journal} {Quantum Studies: Mathematics and Foundations}\ }\textbf {\bibinfo
  {volume} {5}},\ \bibinfo {pages} {579--588} (\bibinfo {year}
  {2017})}\BibitemShut {NoStop}%
\bibitem [{\citenamefont {Jordan}\ \emph {et~al.}(2015)\citenamefont {Jordan},
  \citenamefont {Tollaksen}, \citenamefont {Troupe}, \citenamefont {Dressel},\
  and\ \citenamefont {Aharonov}}]{jordan2015heisenberg}%
  \BibitemOpen
  \bibfield  {author} {\bibinfo {author} {\bibfnamefont {Andrew~N.}\
  \bibnamefont {Jordan}}, \bibinfo {author} {\bibfnamefont {Jeff}\ \bibnamefont
  {Tollaksen}}, \bibinfo {author} {\bibfnamefont {James~E.}\ \bibnamefont
  {Troupe}}, \bibinfo {author} {\bibfnamefont {Justin}\ \bibnamefont
  {Dressel}}, \ and\ \bibinfo {author} {\bibfnamefont {Yakir}\ \bibnamefont
  {Aharonov}},\ }\bibfield  {title} {\enquote {\bibinfo {title} {Heisenberg
  scaling with weak measurement: a quantum state discrimination point of
  view},}\ }\href {\doibase 10.1007/s40509-015-0036-8} {\bibfield  {journal}
  {\bibinfo  {journal} {Quantum Studies: Mathematics and Foundations}\ }\textbf
  {\bibinfo {volume} {2}},\ \bibinfo {pages} {5--15} (\bibinfo {year}
  {2015})}\BibitemShut {NoStop}%
\bibitem [{\citenamefont {Steinberg}(2010)}]{Steinberg10}%
  \BibitemOpen
  \bibfield  {author} {\bibinfo {author} {\bibfnamefont {Aephraim~M.}\
  \bibnamefont {Steinberg}},\ }\bibfield  {title} {\enquote {\bibinfo {title}
  {Quantum measurement: A light touch},}\ }\href
  {https://www.nature.com/articles/463890a} {\bibfield  {journal} {\bibinfo
  {journal} {Nature}\ }\textbf {\bibinfo {volume} {463}},\ \bibinfo {pages}
  {890} (\bibinfo {year} {2010})}\BibitemShut {NoStop}%
\bibitem [{\citenamefont {Wahr}\ \emph {et~al.}()\citenamefont {Wahr},
  \citenamefont {Swenson}, \citenamefont {Zlotnicki},\ and\ \citenamefont
  {Velicogna}}]{Wahr04}%
  \BibitemOpen
  \bibfield  {author} {\bibinfo {author} {\bibfnamefont {John}\ \bibnamefont
  {Wahr}}, \bibinfo {author} {\bibfnamefont {Sean}\ \bibnamefont {Swenson}},
  \bibinfo {author} {\bibfnamefont {Victor}\ \bibnamefont {Zlotnicki}}, \ and\
  \bibinfo {author} {\bibfnamefont {Isabella}\ \bibnamefont {Velicogna}},\
  }\bibfield  {title} {\enquote {\bibinfo {title} {{Time-variable gravity from
  GRACE: First results}},}\ }\href
  {https://agupubs.onlinelibrary.wiley.com/doi/abs/10.1029/2004GL019779}
  {\bibfield  {journal} {\bibinfo  {journal} {Geophysical Research Letters}\
  }\textbf {\bibinfo {volume} {31}}}\BibitemShut {NoStop}%
\bibitem [{\citenamefont {Bingham}\ \emph {et~al.}(2010)\citenamefont
  {Bingham}, \citenamefont {Knudsen}, \citenamefont {Andersen},\ and\
  \citenamefont {Pail}}]{Bingham10}%
  \BibitemOpen
  \bibfield  {author} {\bibinfo {author} {\bibfnamefont {R.~J.}\ \bibnamefont
  {Bingham}}, \bibinfo {author} {\bibfnamefont {P.}~\bibnamefont {Knudsen}},
  \bibinfo {author} {\bibfnamefont {O.~B.}\ \bibnamefont {Andersen}}, \ and\
  \bibinfo {author} {\bibfnamefont {R.}~\bibnamefont {Pail}},\ }\bibfield
  {title} {\enquote {\bibinfo {title} {{Using GOCE to estimate the mean North
  Atlantic circulation}},}\ }in\ \href
  {http://abstractsearch.agu.org/meetings/2010/FM/G33B-08.html} {\emph
  {\bibinfo {booktitle} {AGU Fall Meeting Abstracts}}}\ (\bibinfo {year}
  {2010})\BibitemShut {NoStop}%
\bibitem [{\citenamefont {Bell}\ and\ \citenamefont {Hansen}(1998)}]{Bell98}%
  \BibitemOpen
  \bibfield  {author} {\bibinfo {author} {\bibfnamefont {Robin~E.}\
  \bibnamefont {Bell}}\ and\ \bibinfo {author} {\bibfnamefont {R.~O.}\
  \bibnamefont {Hansen}},\ }\bibfield  {title} {\enquote {\bibinfo {title} {The
  rise and fall of early oil field technology; the torsion balance
  gradiometer},}\ }\href {http://dx.doi.org/} {\bibfield  {journal} {\bibinfo
  {journal} {The Leading Edge}\ }\textbf {\bibinfo {volume} {17}},\ \bibinfo
  {pages} {81} (\bibinfo {year} {1998})}\BibitemShut {NoStop}%
\bibitem [{\citenamefont {van Leeuwen}(2000)}]{Leeuwen00}%
  \BibitemOpen
  \bibfield  {author} {\bibinfo {author} {\bibfnamefont {Edwin~H.}\
  \bibnamefont {van Leeuwen}},\ }\bibfield  {title} {\enquote {\bibinfo {title}
  {{BHP develops airborne gravity gradiometer for mineral exploration}},}\
  }\href {\doibase 10.1190/1.1438526} {\bibfield  {journal} {\bibinfo
  {journal} {The Leading Edge}\ }\textbf {\bibinfo {volume} {19}},\ \bibinfo
  {pages} {1296} (\bibinfo {year} {2000})}\BibitemShut {NoStop}%
\bibitem [{\citenamefont {Diorio}\ \emph {et~al.}(2003)\citenamefont {Diorio},
  \citenamefont {Mahanta}, \citenamefont {Rose},\ and\ \citenamefont
  {Lockhart}}]{Diorio03}%
  \BibitemOpen
  \bibfield  {author} {\bibinfo {author} {\bibfnamefont {P.}~\bibnamefont
  {Diorio}}, \bibinfo {author} {\bibfnamefont {A.}~\bibnamefont {Mahanta}},
  \bibinfo {author} {\bibfnamefont {M.}~\bibnamefont {Rose}}, \ and\ \bibinfo
  {author} {\bibfnamefont {G.}~\bibnamefont {Lockhart}},\ }\bibfield  {title}
  {\enquote {\bibinfo {title} {{Examples of the application of airborne gravity
  gradiometry to natural resource exploration}},}\ }in\ \href@noop {} {\emph
  {\bibinfo {booktitle} {Geophysical Research Abstracts}}},\ Vol.~\bibinfo
  {volume} {5}\ (\bibinfo {year} {2003})\ p.\ \bibinfo {pages}
  {03996}\BibitemShut {NoStop}%
\bibitem [{\citenamefont {Romaides}\ \emph {et~al.}(2001)\citenamefont
  {Romaides}, \citenamefont {Battis}, \citenamefont {Sands}, \citenamefont
  {Zorn}, \citenamefont {Benson~Jr.},\ and\ \citenamefont
  {DiFrancesco}}]{Romaides01}%
  \BibitemOpen
  \bibfield  {author} {\bibinfo {author} {\bibfnamefont {Anestis~J.}\
  \bibnamefont {Romaides}}, \bibinfo {author} {\bibfnamefont {James~C.}\
  \bibnamefont {Battis}}, \bibinfo {author} {\bibfnamefont {Roger~W.}\
  \bibnamefont {Sands}}, \bibinfo {author} {\bibfnamefont {Alan}\ \bibnamefont
  {Zorn}}, \bibinfo {author} {\bibfnamefont {Donald~O.}\ \bibnamefont
  {Benson~Jr.}}, \ and\ \bibinfo {author} {\bibfnamefont {Daniel~J.}\
  \bibnamefont {DiFrancesco}},\ }\bibfield  {title} {\enquote {\bibinfo {title}
  {A comparison of gravimetric techniques for measuring subsurface void
  signals},}\ }\href
  {http://iopscience.iop.org/article/10.1088/0022-3727/34/3/331/meta}
  {\bibfield  {journal} {\bibinfo  {journal} {Journal of Physics D: Applied
  Physics}\ }\textbf {\bibinfo {volume} {34}},\ \bibinfo {pages} {433}
  (\bibinfo {year} {2001})}\BibitemShut {NoStop}%
\bibitem [{\citenamefont {Peters}\ \emph {et~al.}(2001)\citenamefont {Peters},
  \citenamefont {Chung},\ and\ \citenamefont {Chu}}]{Peters01}%
  \BibitemOpen
  \bibfield  {author} {\bibinfo {author} {\bibfnamefont {A.}~\bibnamefont
  {Peters}}, \bibinfo {author} {\bibfnamefont {K.~Y.}\ \bibnamefont {Chung}}, \
  and\ \bibinfo {author} {\bibfnamefont {S.}~\bibnamefont {Chu}},\ }\bibfield
  {title} {\enquote {\bibinfo {title} {High-precision gravity measurements
  using atom interferometry},}\ }\href
  {http://stacks.iop.org/0026-1394/38/i=1/a=4} {\bibfield  {journal} {\bibinfo
  {journal} {Metrologia}\ }\textbf {\bibinfo {volume} {38}},\ \bibinfo {pages}
  {25} (\bibinfo {year} {2001})}\BibitemShut {NoStop}%
\bibitem [{\citenamefont {Luther}\ and\ \citenamefont
  {Towler}(1982)}]{Luther82}%
  \BibitemOpen
  \bibfield  {author} {\bibinfo {author} {\bibfnamefont {Gabriel~G.}\
  \bibnamefont {Luther}}\ and\ \bibinfo {author} {\bibfnamefont {William~R.}\
  \bibnamefont {Towler}},\ }\bibfield  {title} {\enquote {\bibinfo {title}
  {Redetermination of the newtonian gravitational constant $g$},}\ }\href
  {\doibase 10.1103/PhysRevLett.48.121} {\bibfield  {journal} {\bibinfo
  {journal} {Phys. Rev. Lett.}\ }\textbf {\bibinfo {volume} {48}},\ \bibinfo
  {pages} {121--123} (\bibinfo {year} {1982})}\BibitemShut {NoStop}%
\bibitem [{\citenamefont {Kuroda}(1995)}]{Kuroda95}%
  \BibitemOpen
  \bibfield  {author} {\bibinfo {author} {\bibfnamefont {Kazuaki}\ \bibnamefont
  {Kuroda}},\ }\bibfield  {title} {\enquote {\bibinfo {title} {Does the
  time-of-swing method give a correct value of the newtonian gravitational
  constant?}}\ }\href {\doibase 10.1103/PhysRevLett.75.2796} {\bibfield
  {journal} {\bibinfo  {journal} {Phys. Rev. Lett.}\ }\textbf {\bibinfo
  {volume} {75}},\ \bibinfo {pages} {2796--2798} (\bibinfo {year}
  {1995})}\BibitemShut {NoStop}%
\bibitem [{\citenamefont {Karagioz}(1996)}]{Karagioz96}%
  \BibitemOpen
  \bibfield  {author} {\bibinfo {author} {\bibfnamefont {O.~V.}\ \bibnamefont
  {Karagioz}},\ }\bibfield  {title} {\enquote {\bibinfo {title} {{Measurement
  of the gravitational constant with a torsion balance}},}\ }\href
  {https://link.springer.com/article/10.1007/BF02377461} {\bibfield  {journal}
  {\bibinfo  {journal} {Meas. Tech.}\ }\textbf {\bibinfo {volume} {39}},\
  \bibinfo {pages} {979} (\bibinfo {year} {1996})}\BibitemShut {NoStop}%
\bibitem [{\citenamefont {Bagley}\ and\ \citenamefont
  {Luther}(1997)}]{Bagley97}%
  \BibitemOpen
  \bibfield  {author} {\bibinfo {author} {\bibfnamefont {Charles~H.}\
  \bibnamefont {Bagley}}\ and\ \bibinfo {author} {\bibfnamefont {Gabriel~G.}\
  \bibnamefont {Luther}},\ }\bibfield  {title} {\enquote {\bibinfo {title}
  {{Preliminary Results of a Determination of the Newtonian Constant of
  Gravitation: A Test of the Kuroda Hypothesis}},}\ }\href {\doibase
  10.1103/PhysRevLett.78.3047} {\bibfield  {journal} {\bibinfo  {journal}
  {Phys. Rev. Lett.}\ }\textbf {\bibinfo {volume} {78}},\ \bibinfo {pages}
  {3047--3050} (\bibinfo {year} {1997})}\BibitemShut {NoStop}%
\bibitem [{\citenamefont {Gundlach}\ and\ \citenamefont
  {Merkowitz}(2000)}]{Gundlach00}%
  \BibitemOpen
  \bibfield  {author} {\bibinfo {author} {\bibfnamefont {Jens~H.}\ \bibnamefont
  {Gundlach}}\ and\ \bibinfo {author} {\bibfnamefont {Stephen~M.}\ \bibnamefont
  {Merkowitz}},\ }\bibfield  {title} {\enquote {\bibinfo {title} {{Measurement
  of Newton's constant using a torsion balance with angular acceleration
  feedback}},}\ }\href {\doibase 10.1103/PhysRevLett.85.2869} {\bibfield
  {journal} {\bibinfo  {journal} {Phys. Rev. Lett.}\ }\textbf {\bibinfo
  {volume} {85}},\ \bibinfo {pages} {2869--2872} (\bibinfo {year}
  {2000})}\BibitemShut {NoStop}%
\bibitem [{\citenamefont {Quinn}\ \emph {et~al.}(2001)\citenamefont {Quinn},
  \citenamefont {Speake}, \citenamefont {Richman}, \citenamefont {Davis},\ and\
  \citenamefont {Picard}}]{Quinn01}%
  \BibitemOpen
  \bibfield  {author} {\bibinfo {author} {\bibfnamefont {T.~J.}\ \bibnamefont
  {Quinn}}, \bibinfo {author} {\bibfnamefont {C.~C.}\ \bibnamefont {Speake}},
  \bibinfo {author} {\bibfnamefont {S.~J.}\ \bibnamefont {Richman}}, \bibinfo
  {author} {\bibfnamefont {R.~S.}\ \bibnamefont {Davis}}, \ and\ \bibinfo
  {author} {\bibfnamefont {A.}~\bibnamefont {Picard}},\ }\bibfield  {title}
  {\enquote {\bibinfo {title} {A new determination of $\mathit{G}$ using two
  methods},}\ }\href {\doibase 10.1103/PhysRevLett.87.111101} {\bibfield
  {journal} {\bibinfo  {journal} {Phys. Rev. Lett.}\ }\textbf {\bibinfo
  {volume} {87}},\ \bibinfo {pages} {111101} (\bibinfo {year}
  {2001})}\BibitemShut {NoStop}%
\bibitem [{\citenamefont {Armstrong}\ and\ \citenamefont
  {Fitzgerald}(2003)}]{Armstrong03}%
  \BibitemOpen
  \bibfield  {author} {\bibinfo {author} {\bibfnamefont {T.~R.}\ \bibnamefont
  {Armstrong}}\ and\ \bibinfo {author} {\bibfnamefont {M.~P.}\ \bibnamefont
  {Fitzgerald}},\ }\bibfield  {title} {\enquote {\bibinfo {title} {New
  measurements of $g$ using the measurement standards laboratory torsion
  balance},}\ }\href {\doibase 10.1103/PhysRevLett.91.201101} {\bibfield
  {journal} {\bibinfo  {journal} {Phys. Rev. Lett.}\ }\textbf {\bibinfo
  {volume} {91}},\ \bibinfo {pages} {201101} (\bibinfo {year}
  {2003})}\BibitemShut {NoStop}%
\bibitem [{\citenamefont {Kleinevo\ss}\ \emph {et~al.}(1999)\citenamefont
  {Kleinevo\ss}, \citenamefont {Meyer}, \citenamefont {Schumacher},\ and\
  \citenamefont {Hartmann}}]{Kleinevoss99}%
  \BibitemOpen
  \bibfield  {author} {\bibinfo {author} {\bibfnamefont {U.}~\bibnamefont
  {Kleinevo\ss}}, \bibinfo {author} {\bibfnamefont {H.}~\bibnamefont {Meyer}},
  \bibinfo {author} {\bibfnamefont {A.}~\bibnamefont {Schumacher}}, \ and\
  \bibinfo {author} {\bibfnamefont {S.}~\bibnamefont {Hartmann}},\ }\bibfield
  {title} {\enquote {\bibinfo {title} {{Absolute measurement of the Newtonian
  force and a determination of G}},}\ }\href
  {http://stacks.iop.org/0957-0233/10/i=6/a=313} {\bibfield  {journal}
  {\bibinfo  {journal} {Measurement Science and Technology}\ }\textbf {\bibinfo
  {volume} {10}},\ \bibinfo {pages} {492} (\bibinfo {year} {1999})}\BibitemShut
  {NoStop}%
\bibitem [{\citenamefont {Parks}\ and\ \citenamefont {Faller}(2010)}]{Parks10}%
  \BibitemOpen
  \bibfield  {author} {\bibinfo {author} {\bibfnamefont {Harold~V.}\
  \bibnamefont {Parks}}\ and\ \bibinfo {author} {\bibfnamefont {James~E.}\
  \bibnamefont {Faller}},\ }\bibfield  {title} {\enquote {\bibinfo {title}
  {Simple pendulum determination of the gravitational constant},}\ }\href
  {\doibase 10.1103/PhysRevLett.105.110801} {\bibfield  {journal} {\bibinfo
  {journal} {Phys. Rev. Lett.}\ }\textbf {\bibinfo {volume} {105}},\ \bibinfo
  {pages} {110801} (\bibinfo {year} {2010})}\BibitemShut {NoStop}%
\bibitem [{\citenamefont {Peters}\ \emph {et~al.}(1999)\citenamefont {Peters},
  \citenamefont {Chung},\ and\ \citenamefont {Chu}}]{Peters99}%
  \BibitemOpen
  \bibfield  {author} {\bibinfo {author} {\bibfnamefont {Achim}\ \bibnamefont
  {Peters}}, \bibinfo {author} {\bibfnamefont {Keng~Yeow}\ \bibnamefont
  {Chung}}, \ and\ \bibinfo {author} {\bibfnamefont {Steven}\ \bibnamefont
  {Chu}},\ }\bibfield  {title} {\enquote {\bibinfo {title} {Measurement of
  gravitational acceleration by dropping atoms},}\ }\href
  {https://www.nature.com/articles/23655} {\bibfield  {journal} {\bibinfo
  {journal} {Nature}\ }\textbf {\bibinfo {volume} {400}},\ \bibinfo {pages}
  {849} (\bibinfo {year} {1999})}\BibitemShut {NoStop}%
\bibitem [{\citenamefont {McGuirk}\ \emph {et~al.}(2002)\citenamefont
  {McGuirk}, \citenamefont {Foster}, \citenamefont {Fixler}, \citenamefont
  {Snadden},\ and\ \citenamefont {Kasevich}}]{Mcguirk02}%
  \BibitemOpen
  \bibfield  {author} {\bibinfo {author} {\bibfnamefont {J.~M.}\ \bibnamefont
  {McGuirk}}, \bibinfo {author} {\bibfnamefont {G.~T.}\ \bibnamefont {Foster}},
  \bibinfo {author} {\bibfnamefont {J.~B.}\ \bibnamefont {Fixler}}, \bibinfo
  {author} {\bibfnamefont {M.~J.}\ \bibnamefont {Snadden}}, \ and\ \bibinfo
  {author} {\bibfnamefont {M.~A.}\ \bibnamefont {Kasevich}},\ }\bibfield
  {title} {\enquote {\bibinfo {title} {Sensitive absolute-gravity gradiometry
  using atom interferometry},}\ }\href {\doibase 10.1103/PhysRevA.65.033608}
  {\bibfield  {journal} {\bibinfo  {journal} {Phys. Rev. A}\ }\textbf {\bibinfo
  {volume} {65}},\ \bibinfo {pages} {033608} (\bibinfo {year}
  {2002})}\BibitemShut {NoStop}%
\bibitem [{\citenamefont {Dimopoulos}\ \emph {et~al.}(2007)\citenamefont
  {Dimopoulos}, \citenamefont {Graham}, \citenamefont {Hogan},\ and\
  \citenamefont {Kasevich}}]{Dimopoulos07}%
  \BibitemOpen
  \bibfield  {author} {\bibinfo {author} {\bibfnamefont {Savas}\ \bibnamefont
  {Dimopoulos}}, \bibinfo {author} {\bibfnamefont {Peter~W.}\ \bibnamefont
  {Graham}}, \bibinfo {author} {\bibfnamefont {Jason~M.}\ \bibnamefont
  {Hogan}}, \ and\ \bibinfo {author} {\bibfnamefont {Mark~A.}\ \bibnamefont
  {Kasevich}},\ }\bibfield  {title} {\enquote {\bibinfo {title} {Testing
  general relativity with atom interferometry},}\ }\href {\doibase
  10.1103/PhysRevLett.98.111102} {\bibfield  {journal} {\bibinfo  {journal}
  {Phys. Rev. Lett.}\ }\textbf {\bibinfo {volume} {98}},\ \bibinfo {pages}
  {111102} (\bibinfo {year} {2007})}\BibitemShut {NoStop}%
\bibitem [{\citenamefont {Lamporesi}\ \emph {et~al.}(2008)\citenamefont
  {Lamporesi}, \citenamefont {Bertoldi}, \citenamefont {Cacciapuoti},
  \citenamefont {Prevedelli},\ and\ \citenamefont {Tino}}]{Lamporesi08}%
  \BibitemOpen
  \bibfield  {author} {\bibinfo {author} {\bibfnamefont {G.}~\bibnamefont
  {Lamporesi}}, \bibinfo {author} {\bibfnamefont {A.}~\bibnamefont {Bertoldi}},
  \bibinfo {author} {\bibfnamefont {L.}~\bibnamefont {Cacciapuoti}}, \bibinfo
  {author} {\bibfnamefont {M.}~\bibnamefont {Prevedelli}}, \ and\ \bibinfo
  {author} {\bibfnamefont {G.~M.}\ \bibnamefont {Tino}},\ }\bibfield  {title}
  {\enquote {\bibinfo {title} {{Determination of the Newtonian gravitational
  constant using atom interferometry}},}\ }\href {\doibase
  10.1103/PhysRevLett.100.050801} {\bibfield  {journal} {\bibinfo  {journal}
  {Phys. Rev. Lett.}\ }\textbf {\bibinfo {volume} {100}},\ \bibinfo {pages}
  {050801} (\bibinfo {year} {2008})}\BibitemShut {NoStop}%
\bibitem [{\citenamefont {Sorrentino}\ \emph {et~al.}(2010)\citenamefont
  {Sorrentino}, \citenamefont {Lien}, \citenamefont {Rosi}, \citenamefont
  {Cacciapuoti}, \citenamefont {Prevedelli},\ and\ \citenamefont
  {Tino}}]{Sorrentino10}%
  \BibitemOpen
  \bibfield  {author} {\bibinfo {author} {\bibfnamefont {F.}~\bibnamefont
  {Sorrentino}}, \bibinfo {author} {\bibfnamefont {Y.~H.}\ \bibnamefont
  {Lien}}, \bibinfo {author} {\bibfnamefont {G.}~\bibnamefont {Rosi}}, \bibinfo
  {author} {\bibfnamefont {L.}~\bibnamefont {Cacciapuoti}}, \bibinfo {author}
  {\bibfnamefont {M.}~\bibnamefont {Prevedelli}}, \ and\ \bibinfo {author}
  {\bibfnamefont {G.~M.}\ \bibnamefont {Tino}},\ }\bibfield  {title} {\enquote
  {\bibinfo {title} {Sensitive gravity-gradiometry with atom interferometry:
  progress towards an improved determination of the gravitational constant},}\
  }\href {http://iopscience.iop.org/article/10.1088/1367-2630/12/9/095009/meta}
  {\bibfield  {journal} {\bibinfo  {journal} {New Journal of Physics}\ }\textbf
  {\bibinfo {volume} {12}},\ \bibinfo {pages} {095009} (\bibinfo {year}
  {2010})}\BibitemShut {NoStop}%
\bibitem [{\citenamefont {Rosi}\ \emph {et~al.}(2014)\citenamefont {Rosi},
  \citenamefont {Sorrentino}, \citenamefont {Cacciapuoti}, \citenamefont
  {Prevedelli},\ and\ \citenamefont {Tino}}]{Rosi14}%
  \BibitemOpen
  \bibfield  {author} {\bibinfo {author} {\bibfnamefont {G.}~\bibnamefont
  {Rosi}}, \bibinfo {author} {\bibfnamefont {F.}~\bibnamefont {Sorrentino}},
  \bibinfo {author} {\bibfnamefont {L.}~\bibnamefont {Cacciapuoti}}, \bibinfo
  {author} {\bibfnamefont {M.}~\bibnamefont {Prevedelli}}, \ and\ \bibinfo
  {author} {\bibfnamefont {G.~M.}\ \bibnamefont {Tino}},\ }\bibfield  {title}
  {\enquote {\bibinfo {title} {{Precision measurement of the Newtonian
  gravitational constant using cold atoms}},}\ }\href
  {https://www.nature.com/articles/nature13433} {\bibfield  {journal} {\bibinfo
   {journal} {Nature}\ }\textbf {\bibinfo {volume} {510}},\ \bibinfo {pages}
  {518} (\bibinfo {year} {2014})}\BibitemShut {NoStop}%
\bibitem [{\citenamefont {Goodkind}(1999)}]{Goodkind99}%
  \BibitemOpen
  \bibfield  {author} {\bibinfo {author} {\bibfnamefont {John~M.}\ \bibnamefont
  {Goodkind}},\ }\bibfield  {title} {\enquote {\bibinfo {title} {The
  superconducting gravimeter},}\ }\href {\doibase 10.1063/1.1150092} {\bibfield
   {journal} {\bibinfo  {journal} {Review of Scientific Instruments}\ }\textbf
  {\bibinfo {volume} {70}},\ \bibinfo {pages} {4131--4152} (\bibinfo {year}
  {1999})}\BibitemShut {NoStop}%
\bibitem [{\citenamefont {Biedermann}\ \emph {et~al.}(2015)\citenamefont
  {Biedermann}, \citenamefont {Wu}, \citenamefont {Deslauriers}, \citenamefont
  {Roy}, \citenamefont {Mahadeswaraswamy},\ and\ \citenamefont
  {Kasevich}}]{PhysRevA.91.033629}%
  \BibitemOpen
  \bibfield  {author} {\bibinfo {author} {\bibfnamefont {G.~W.}\ \bibnamefont
  {Biedermann}}, \bibinfo {author} {\bibfnamefont {X.}~\bibnamefont {Wu}},
  \bibinfo {author} {\bibfnamefont {L.}~\bibnamefont {Deslauriers}}, \bibinfo
  {author} {\bibfnamefont {S.}~\bibnamefont {Roy}}, \bibinfo {author}
  {\bibfnamefont {C.}~\bibnamefont {Mahadeswaraswamy}}, \ and\ \bibinfo
  {author} {\bibfnamefont {M.~A.}\ \bibnamefont {Kasevich}},\ }\bibfield
  {title} {\enquote {\bibinfo {title} {Testing gravity with cold-atom
  interferometers},}\ }\href {\doibase 10.1103/PhysRevA.91.033629} {\bibfield
  {journal} {\bibinfo  {journal} {Phys. Rev. A}\ }\textbf {\bibinfo {volume}
  {91}},\ \bibinfo {pages} {033629} (\bibinfo {year} {2015})}\BibitemShut
  {NoStop}%
\bibitem [{\citenamefont {Kasevich}\ \emph {et~al.}(2014)\citenamefont
  {Kasevich}, \citenamefont {Donnelly},\ and\ \citenamefont
  {Overstreet}}]{kasevich2014prospects}%
  \BibitemOpen
  \bibfield  {author} {\bibinfo {author} {\bibfnamefont {Mark}\ \bibnamefont
  {Kasevich}}, \bibinfo {author} {\bibfnamefont {Christine}\ \bibnamefont
  {Donnelly}}, \ and\ \bibinfo {author} {\bibfnamefont {Chris}\ \bibnamefont
  {Overstreet}},\ }\bibfield  {title} {\enquote {\bibinfo {title} {{Prospects
  for improved accuracy in the determination of G using atom
  interferometry}},}\ }\href
  {https://pdfs.semanticscholar.org/presentation/3a8d/94dcb6fcd7c381e2b51787d7e5e39438a6c5.pdf}
  {\bibfield  {journal} {\bibinfo  {journal} {Depts. of Physics, Applied
  Physics and EE Stanford University}\ } (\bibinfo {year} {2014})}\BibitemShut
  {NoStop}%
\bibitem [{\citenamefont {Turner}(2018)}]{turner2018development}%
  \BibitemOpen
  \bibfield  {author} {\bibinfo {author} {\bibfnamefont {Matthew~David}\
  \bibnamefont {Turner}},\ }\emph {\bibinfo {title} {Development of new
  technologies for precision torsion-balance experiments}},\ \href
  {https://digital.lib.washington.edu/researchworks/handle/1773/40958?show=full}
  {Ph.D. thesis} (\bibinfo {year} {2018})\BibitemShut {NoStop}%
\bibitem [{\citenamefont {Ciani}\ \emph {et~al.}(2017)\citenamefont {Ciani},
  \citenamefont {Chilton}, \citenamefont {Apple}, \citenamefont {Olatunde},
  \citenamefont {Aitken}, \citenamefont {Mueller},\ and\ \citenamefont
  {Conklin}}]{Ciani2017}%
  \BibitemOpen
  \bibfield  {author} {\bibinfo {author} {\bibfnamefont {Giacomo}\ \bibnamefont
  {Ciani}}, \bibinfo {author} {\bibfnamefont {Andrew}\ \bibnamefont {Chilton}},
  \bibinfo {author} {\bibfnamefont {Stephen}\ \bibnamefont {Apple}}, \bibinfo
  {author} {\bibfnamefont {Taiwo}\ \bibnamefont {Olatunde}}, \bibinfo {author}
  {\bibfnamefont {Michael}\ \bibnamefont {Aitken}}, \bibinfo {author}
  {\bibfnamefont {Guido}\ \bibnamefont {Mueller}}, \ and\ \bibinfo {author}
  {\bibfnamefont {John~W.}\ \bibnamefont {Conklin}},\ }\bibfield  {title}
  {\enquote {\bibinfo {title} {A new torsion pendulum for gravitational
  reference sensor technology development},}\ }\href {\doibase
  10.1063/1.4985543} {\bibfield  {journal} {\bibinfo  {journal} {Review of
  Scientific Instruments}\ }\textbf {\bibinfo {volume} {88}},\ \bibinfo {pages}
  {064502} (\bibinfo {year} {2017})}\BibitemShut {NoStop}%
\bibitem [{\citenamefont {Starling}\ \emph {et~al.}(2010)\citenamefont
  {Starling}, \citenamefont {Dixon}, \citenamefont {Williams}, \citenamefont
  {Jordan},\ and\ \citenamefont {Howell}}]{starling2010continuous}%
  \BibitemOpen
  \bibfield  {author} {\bibinfo {author} {\bibfnamefont {David~J.}\
  \bibnamefont {Starling}}, \bibinfo {author} {\bibfnamefont {P.~Ben}\
  \bibnamefont {Dixon}}, \bibinfo {author} {\bibfnamefont {Nathan~S.}\
  \bibnamefont {Williams}}, \bibinfo {author} {\bibfnamefont {Andrew~N.}\
  \bibnamefont {Jordan}}, \ and\ \bibinfo {author} {\bibfnamefont {John~C.}\
  \bibnamefont {Howell}},\ }\bibfield  {title} {\enquote {\bibinfo {title}
  {{Continuous phase amplification with a Sagnac interferometer}},}\ }\href
  {\doibase 10.1103/PhysRevA.82.011802} {\bibfield  {journal} {\bibinfo
  {journal} {Phys. Rev. A}\ }\textbf {\bibinfo {volume} {82}},\ \bibinfo
  {pages} {011802} (\bibinfo {year} {2010})}\BibitemShut {NoStop}%
\bibitem [{\citenamefont {Knee}\ and\ \citenamefont
  {Gauger}(2014)}]{knee2014amplification}%
  \BibitemOpen
  \bibfield  {author} {\bibinfo {author} {\bibfnamefont {George~C.}\
  \bibnamefont {Knee}}\ and\ \bibinfo {author} {\bibfnamefont {Erik~M.}\
  \bibnamefont {Gauger}},\ }\bibfield  {title} {\enquote {\bibinfo {title}
  {When amplification with weak values fails to suppress technical noise},}\
  }\href {\doibase 10.1103/PhysRevX.4.011032} {\bibfield  {journal} {\bibinfo
  {journal} {Phys. Rev. X}\ }\textbf {\bibinfo {volume} {4}},\ \bibinfo {pages}
  {011032} (\bibinfo {year} {2014})}\BibitemShut {NoStop}%
\bibitem [{\citenamefont {Newton}(1687)}]{newton1687}%
  \BibitemOpen
  \bibfield  {author} {\bibinfo {author} {\bibfnamefont {I.}~\bibnamefont
  {Newton}},\ }\href {https://books.google.com/books?id=-dVKAQAAIAAJ} {\emph
  {\bibinfo {title} {Philosophiae naturalis principia mathematica}}}\ (\bibinfo
   {publisher} {J. Societatis Regiae ac Typis J. Streater},\ \bibinfo {year}
  {1687})\BibitemShut {NoStop}%
\end{thebibliography}%

\end{document}